\documentclass[12pt]{article}

\usepackage[letterpaper,top=2.54cm,bottom=2.54cm,left=2.54cm,right=2.54cm]{geometry}
\usepackage{setspace}
\usepackage[english]{babel}
\usepackage{enumitem}
\usepackage[normalem]{ulem}
\usepackage{xcolor}
\usepackage{soul}
\usepackage[round]{natbib}
\usepackage{hyperref}

\usepackage{graphicx} 
\usepackage{placeins}
\usepackage{amsmath}
\usepackage{amssymb, amsfonts, bm}

\title{TRAECR: A Tool for Preprocessing Positron Emission Tomography Imaging for Statistical Modeling}

\author{Akhil Ambekar,
Robert Zielinski,
and 
Ani Eloyan\\
   Center for Biostatistics and Health Data Science, Department of Biostatistics, \\ Brown University, Providence, RI, USA\\
}

\date{\today}

\begin{document}
\doublespacing

\maketitle

\begin{abstract}
    Positron emission tomography (PET) imaging is widely used in a number of clinical applications, including cancer and Alzheimer's disease (AD) diagnosis, monitoring of disease development, and treatment effect evaluation. Statistical modeling of PET imaging is essential to address continually emerging scientific questions in these research fields, including hypotheses related to evaluation of effects of disease modifying treatments on amyloid reduction in AD and associations between amyloid reduction and cognitive function, among many others. In this paper, we provide background information and tools for statisticians interested in developing statistical models for PET imaging to pre-process and prepare data for analysis. We introduce our novel pre-processing and visualization tool TRAECR (Template registration, MRI-PET co-Registration, Anatomical brain Extraction and COMBAT/RAVEL harmonization) to facilitate data preparation for statistical analysis.  
\end{abstract}


\section{Introduction}

Brain imaging technologies, such as Magnetic Resonance Imaging (MRI) and Positron Emission Tomography (PET), include important tools widely used in modern medical research for clinical diagnosis of neurodegerative diseases and monitoring of disease progression. These modalities provide detailed visualizations of the brain's anatomical structures, evaluation of functional activity in the brain, and crucial information on glucose metabolism in specific brain regions. These findings are essential for understanding neurological disorders, guiding surgical procedures, and monitoring response to therapeutic interventions. The precision and reliability of the images collected by these technologies as well as the modeling pipelines implemented to analyze the data are critical, as they directly influence diagnostic decisions and subsequent medical or research outcomes \citep{schwarz2021uses}. \cite{jones2017history} provide a thorough overview of the history of innovations that lead to the introduction of PET imaging and developments of this technology since its inception. 

Pre-processing pipelines are an important step for preparation of images for statistical modeling. When modeling imaging data to address population level hypotheses, statisticians often work with data from large, multicenter studies, where data is collected in different sites. Pre-processing brain imaging data from modalities like MRI and PET presents various challenges that can impact data quality and utility. Key sources of systematic noise impacting image quality in most multi-center imaging studies include differences in scanner hardware (e.g. manufacturer, magnetic field strength), software protocols (e.g. pulse sequences, reconstruction algorithms), and image acquisition parameters (e.g. voxel size, echo time, repetition time) \citep{hagiwara2020variability}. These variations can lead to significant discrepancies in image quality and characteristics both between and within data collection sites. This variability between scanners complicates comparisons between study populations and can introduce biases in multicenter research \citep{thieleking2021same}. In addition, MRI and PET scans are prone to various artifacts, including motion-blur, susceptibility distortions, and noise arising from patient movement or physiological processes. These artifacts can obscure critical details and introduce bias in image analysis results \citep{vandenberghe2015pet}. Accurate alignment and registration of brain images, whether within participant or to a common template space, are essential in multi-modal studies or longitudinal monitoring, requiring sophisticated computational techniques to ensure a precise overlay of anatomical and functional data \citep{deng2024deep}. Due to these pre-processing challenges, solutions that can standardize, enhance, and harmonize images before performing analysis are essential. Since ``concurrent'' MRI scans are often used for pre-processing of PET images, we discuss MRI and PET pre-processing together henceforth. 

The R programming language \citep{r2013r} is increasingly used for modeling MRI and fMRI data with the introduction of various packages for processing and modeling imaging data available in public software repositories, including Comprehensive R Archive Network (CRAN) and neuroconductor \citep{muschelli2019neuroconductor}. Existing pre-processing software suites with interfaces in R include the FMRIB Software Library (FSL) \citep{smith2004advances}, Statistical Parametric Mapping (SPM) \citep{penny2011statistical}, and Advanced Normalization Tools (ANTs) \citep{avants2011reproducible}, which offer pre-processing functionalities such as skull stripping, motion correction, artifact reduction, and image registration. However, these solutions often require extensive tuning and coding expertise, as well as familiarity with the individual packages used in the pre-processing steps. Integrating multiple pre-processing steps across different software packages can introduce complexity and potential errors in data handling, increasing the burden on researchers and clinicians. In addition, pre-processing PET data, while often including steps similar to those used for pre-processing other data modalities (e.g. registration to a common template space, skull stripping), has specific differences. Hence, it is important to consider the adaptation of pre-processing pipelines when modeling PET images. These factors make it difficult for researchers to comprehensively address all needs within a unified workflow. These limitations may result in barriers for researchers new to the field of PET imaging analysis.

To mitigate these challenges, we developed TRAECR, an integrated PET pre-processing tool designed with a user-friendly interface and comprehensive functionality, implemented as an R Shiny application. TRAECR includes an artifact-detection dashboard developed in Python \citep{rossum1995python} using Plotly’s Dash framework \citep{sievert2020interactive}. This dashboard is designed to support MRI and PET quality control (QC) by enabling interactive image review, visualization of quantitative image quality metrics and robust z-scores, and inspection/export of available image metadata.  
It provides a useful mechanism for determining which image files should be included for pre-processing or excluded from further analysis. 

TRAECR is built upon established tools available in existing neuroimaging software and is intended to integrate these components into a streamlined preprocessing workflow. The core features of TRAECR include brain extraction; template registration; MRI-PET co-registration; corrections for batch effects and scanner variability; and adjustments for variability caused by non-biological factors. Brain extraction is performed using the \texttt{fslbet\_robust} function from \texttt{extrantsr} package in R. This approach efficiently isolates brain tissue from MRI scans by removing non-brain structures, including the scalp, skull, and neck, which helps produce brain-only images. The tool also registers the brain extracted images to the Montreal Neurological Institute (MNI) template space \citep{mazziotta1995probabilistic} using the FMRIB's Linear Image Registration Tool (FLIRT) algorithm \citep{smith2004advances}, facilitating standardized comparisons across participants and studies. Next, the MRI-PET co-registration functionality of TRAECR aligns PET images with corresponding MRI scans to enable precise overlay of functional data onto anatomical structures, enhancing multi-modal analyses. 
For both template registration and MRI--PET co-registration, an optional atlas based parcellation feature is provided that resamples atlas labels with nearest neighbor interpolation and outputs a label map, individual ROI masks, and ROI wise mean intensity values.
To adjust for batch effects and scanner induced variability, the tool implements the COMBAT harmonization method \citep{fortin2018harmonization} that improves data comparability between different scanners or study sites. Finally, TRAECR utilizes the RAVEL normalization algorithm \citep{fortin2016removing} to reduce variability arising from non-biological factors, such as scanner related differences, ensuring that true biological signals are preserved in the data.

By integrating these functionalities into a single, user-friendly platform, TRAECR simplifies the pre-processing workflow while improving the consistency and comparability of brain imaging data across different studies and scanners. It offers researchers and clinicians a streamlined approach to handle essential pre-processing, enhancing efficiency, reducing errors, and saving valuable time and effort in neuroimaging statistics research. This paper serves as both a tutorial and an introduction to a useful tool for pre-processing for statisticians new to modeling PET imaging data. 

The manuscript is organized as follows. In Section \ref{s:background}, we provide background on PET imaging and quantitative analysis approaches relevant to downstream statistical modeling. We then describe the TRAECR quality-control dashboard, image-quality metrics, metadata viewer, and core preprocessing modules in Section \ref{s:methods}. Next, in Section \ref{s:Algorithms} we outline the algorithms implemented for brain extraction, template registration, MRI–PET co-registration, atlas-based parcellation, COMBAT harmonization, and RAVEL normalization. Finally, we describe the computational setup and deployment workflow in Section \ref{s:toolsetup}. We conclude the manuscript with discussion of limitations, and planned future extensions in Section \ref{s:conclusions}.

\section{Background}\label{s:background}

PET has become a crucial imaging technique in modern medicine, providing deep insight into the physiological and molecular functions of the human body. By using radiotracers tagged with positron emitting isotopes, PET enables minimally invasive visualization and measurement of biochemical processes in living organism (\textit{in vivo})  \citep{lameka2016positron}. This powerful capability is important for the early detection of diseases such as Alzheimer's disease (AD), tracking how patients respond to treatments, and enhancing understanding of various pathological conditions. In this section, we provide a broad overview of common uses of PET imaging data in clinical and research applications. 

In oncology, PET imaging has transformed cancer diagnosis and management for the better, particularly utilizing [\textsuperscript{18}F]-fluorodeoxyglucose (FDG). FDG-PET leverages the heightened glucose metabolism inherent to malignant cells, enabling the detection and staging of cancers, including lymphoma, lung, colorectal, and breast cancers \citep{boellaard2015fdg}. This imaging modality provides critical insights on tumor metabolism, facilitating the differentiation between benign and malignant lesions, evaluating treatment response, and early detection of disease recurrence \citep{wahl2009recist}. Additionally, the quantitative evaluation of standardized uptake values (SUVs) in PET imaging is essential for measuring metabolic activity and monitoring temporal changes, thus supporting informed clinical decision making \citep{boellaard2015fdg}.

In neurology, PET imaging has advanced understanding and diagnosis of neurodegenerative diseases such as AD. PET tracers targeting amyloid-beta (A$\beta$) plaques and tau proteins, such as [\textsuperscript{11}C]-PiB and [\textsuperscript{18}F]-flortaucipir, enable early detection and diagnosis of AD by visualizing these pathological hallmarks before clinical symptoms become apparent. Amyloid PET quantifies the burden and spatial distribution of A$\beta$ deposition, providing biomarkers useful for disease characterization and for differentiating AD from other dementias. By contrast, tau PET shows stronger regional associations with neurodegeneration and clinical heterogeneity, offering complementary insight into disease progression \citep{ossenkoppele2016tau}. Furthermore, PET imaging of dopaminergic function using [\textsuperscript{18}F]-DOPA assists in diagnosing Parkinson's disease and other movement disorders by assessing the integrity of the nigrostriatal pathway \citep{brooks2010imaging}. This application is essential for differentiating Parkinsonian syndromes from other neurological conditions with similar clinical presentations. The ability to visualize and quantify these molecular targets {\it in vivo} highlights an important role of PET imaging in both clinical diagnostics and the development of targeted therapeutic strategies for neurodegenerative disorders.

PET has also been useful in other fields of medicine. In cardiology, PET imaging is used for assessment of myocardial perfusion and viability. PET tracers like [\textsuperscript{13}N]-ammonia and [\textsuperscript{82}Rb]-rubidium are used to evaluate myocardial blood flow, aiding in the diagnosis of coronary artery disease \citep{tsj2018myocardial}. PET imaging can detect myocardial tissue that is still alive, but suffers from reduced blood supply, guiding revascularization strategies and improving patient outcomes. Beyond clinical applications, PET is a powerful tool in drug development and pharmacokinetics. By labeling pharmaceutical compounds with positron emitting isotopes, PET allows for the \textit{in vivo} tracking of drug distribution, receptor occupancy, and metabolism. This information accelerates the drug development process by providing early insights into pharmacodynamics and optimal dosing regimens \citep{weissleder2006molecular}.

Reliable post-reconstruction pre-processing is important for ensuring consistency in population level PET analysis and for supporting downstream quantitative modeling. Although PET image quality can be affected by factors such as photon attenuation, scatter, random coincidences, and patient motion \citep{berker2016attenuation,frey2012accuracy,rahmim2008pet}, these effects are typically addressed during acquisition and scanner reconstruction. In this setting, TRAECR is intended for reconstructed PET images, or pre-averaged images derived from dynamic acquisitions, along with corresponding MRI data, and provides tools for downstream pre-processing and quality control. In addition, TRAECR can be used for pre-processing PET data collected for different radiotracers. The selection of specific tracer(s) for analysis is determined by the user. However, when multiple tracers are included, harmonization and normalization methods, such as COMBAT and RAVEL, should be applied separately within each tracer type rather than across all tracers combined, because tracer specific uptake patterns and intensity distributions may differ substantially.

\subsection{Statistical Analyses of PET Imaging Data} 

In this Section, we briefly discuss a few statistical modeling approaches in PET imaging studies. \citep{nichols2001spatiotemporal} provides perspectives on spatiotemporal modeling of PET, and \citep{ombao2016handbook} discuss the properties of PET and common statistical modeling (see Chapter 2 and references therein). Depending on the goals of the study, either static or dynamic PET data are typically analyzed. Static PET acquisition allows for a fast data collection to investigate spatial patterns of ligand accumulation in the organ of interest. In static PET the standardized uptake value (SUV) is a commonly used measurement of mean activity concentration in a given ROI, which is then normalized by the body weight and the dose of the injected tracer. In contrast, dynamic PET is collected over several time points after injection of the tracer. These data can be used to evaluate the rates of movement between blood and compartments of the tissue. Discussion of compartmental models for addressing many hypotheses in PET imaging analysis along with a survey of literature on this topic is described in Chapter 2 of \citep{ombao2016handbook}. 

After MRI–PET co-registration and inter-scanner harmonization with COMBAT and RAVEL, regional tracer uptake is quantified via the standardized uptake value ratio (SUVR). The SUVR converts a static PET image into a dimensionless map that can be compared across subjects and scanners \citep{boellaard2009standards}.  First, at time $t$ post-injection, the voxel-wise standardized uptake value is obtained as  
\[
\mathrm{SUV}(t)=\frac{C_{\text{tissue}}(t)\,[\mathrm{MBq}\,\mathrm{ml}^{-1}]}{D_{\text{inj}}(\mathrm{MBq})\,/\,m_{\text{body}}}\; ,
\]
where \(C_{\text{tissue}}(t)\) is the decay-corrected activity concentration at time \(t\), \(D_{\text{inj}}\) (MBq) is the injected dose and \(m_{\text{body}}\) is body mass (or lean-body mass/body-surface area, depending on the study protocol).  
The SUVR in a target region \(T\) is then  
\[
\mathrm{SUVR}_{T/R}= \frac{\overline{\mathrm{SUV}}_{T}}{\overline{\mathrm{SUV}}_{R}},
\]
with \(R\) a reference region assumed to have negligible specific binding, and the overline denoting the spatial (voxel-wise) arithmetic mean within the region of interest.

Because SUVR relies on images taken when tracer uptake is nearly constant, data should be acquired during the tracer’s near-equilibrium phase.  For amyloid tracers such as \({}^{11}\text{C}\)-PiB this phase is typically 50–70 min after injection, while for \({}^{18}\text{F}\)-FDG it is about 30–60 min \citep{mcnamee2009consideration}.  Using one fixed static frame in these windows limits variability from blood-flow differences and removes the need for invasive arterial sampling needed for many dynamic PET parameter estimation procedures. The choice of the reference region \(R\) has a major impact on SUVR precision and longitudinal stability. This choice is dependent on the PET tracer or the hypothesis of interest in each study. For example, in \emph{Amyloid PET} whole cerebellar or cerebellar-cortex uptake is widely used because amyloid plaques are essentially absent in these regions even in advanced Alzheimer’s disease (AD).  In \emph{tau PET} for cross-sectional studies the inferior cerebellar gray matter is preferred; for longitudinal studies an eroded white-matter composite or brain-stem/partial-volume-corrected cerebellum can further reduce variance. Finally, in \emph{oncology or whole body PET,} the liver or descending aorta often serves as an internal reference to account for patient-specific systemic factors. 

SUVR has limitations as it is sensitive to cerebral blood-flow changes, susceptible to spill-in/out from off-target binding, and its variance inflates when reference-region noise is high \citep{ottoy2017simulation}.  Nevertheless, when acquisition, reference region and processing are standardized, SUVR offers a rapid, low-burden surrogate for full kinetic modeling and is the dominant endpoint in large-scale population and therapeutic PET studies.

A typical SUVR workflow includes the following steps:

\begin{enumerate}
  \item Motion correct the dynamic or list-mode data; reconstruct the static frame (e.g.\ 50–70 min).  
  \item Co-register PET to the subject’s MRI and segment cortical/subcortical ROIs (FreeSurfer, SPM, or similar).  
  \item Apply scanner specific resolution normalization (or partial volume correction) to reduce apparent uptake loss in atrophied cortex.  
  \item Compute mean SUV in each ROI and divide by the reference-region SUV to obtain regional SUVRs; optionally project voxel-wise SUVR maps into template space for statistical parametric mapping.  
\end{enumerate}

We designed TRAECR to simplify key parts of this workflow.

\section{Methods}\label{s:methods}

In this section, we describe the TRAECR architecture, the interactive dashboard for artifact detection, the quantitative image quality metrics it computes, and the suite of core pre-processing utilities that together make up the tool.

\subsection{Integrated Dashboard for Artifact Detection}

To enable rapid first-pass QC, we developed an interactive image properties dashboard within TRAECR for screening and flagging potentially bad quality MRI and PET scans using distributional summaries of voxel intensities. The {\it gold standard} quality control for PET imaging remains the visual read by a trained radiologist. However, software for quality control for MRI has been developed to compute metrics for identifying imaging with possible quality issues for further inspection such as the MRIQC \citep{esteban2017mriqc}. Following this established approach for MRI, we implemented a quality control approach in Python using the Dash framework. The dashboard provides an easy to use web interface for batch upload of Neuroimaging Informatics Technology Initiative (NIfTI) images. Upon upload, it automatically computes quantitative metrics for each scan and highlights outliers to prioritize targeted visual inspection rather than manual scan by scan review. While the user may choose to visually inspect each image in their collection, this can be quite time consuming in large imaging studies that are often of interest for statistical modeling. Our approach provides a list of potential outliers in terms of their intensity histogram characteristics, useful for large datasets where individual image inspection may not be feasible.

A key design choice is that each metric is computed within the whole-brain, defined here as all non-zero voxels in the input image. After loading the image, all summary metrics, robust z-scores, and interpretation are based on the non-zero voxels. Details of the procedure are described next.

\subsubsection{Image Quality Metrics}

The dashboard computes summary statistics from voxel intensity distributions and uses them as image quality metrics to flag scans that may warrant closer review. Specifically, the dashboard reports the mean, median, standard deviation, variance, percentiles (10th, 90th, and 99th), and root mean square (RMS) of the whole-brain voxel intensity histogram, as well as its dispersion and histogram shape descriptors including mean absolute deviation about the mean (\texttt{mad\_mean}), median absolute deviation about the median (\texttt{mad\_median}), Shannon entropy, uniformity, skewness, and kurtosis. For example, elevated entropy may reflect increased noise, atypical intensity heterogeneity, or potential normalization/rescaling issues, whereas very low entropy may suggest an unusually uniform image, a near-empty image, heavy clipping, or reduced dynamic range. In addition, several QC-oriented diagnostic metrics are included: a proxy signal-to-noise ratio (SNR), defined here as the mean intensity divided by the standard deviation; coefficient of variation (CV), defined as the standard deviation divided by the mean intensity; intensity range; and clipping fraction, defined as the fraction of voxels at the maximum intensity. A high clipping fraction may indicate saturation, intensity clipping, or rescaling artifacts, although these metrics should be interpreted as screening indicators rather than definitive failure criteria. The \textit{Interpretation \& Guidelines} tab in the dashboard summarizes these potential interpretations and provides practical guidance for reviewing flagged scans, applying robust-z thresholds, and using the reported metrics during manual follow-up (Supplementary Figure 3).

Because MRI and PET intensity scales vary substantially across scanners, acquisition protocols, tracers, reconstruction settings, and pre-processing pipelines, absolute cutoffs for histogram based metrics are not reliable across studies. The dashboard therefore emphasizes within batch robust outlier detection by computing a robust z-score for each metric using the batch median and median absolute deviation (MAD). Rather than enforcing fixed universal thresholds, the dashboard allows the user to select the z-score cutoff interactively for their specific dataset and review goals. In the current implementation, values $|z| \ge 3$ for \textbf{WARN} and \textbf{$|z| \ge 5$} for \textbf{FAIL} are provided as example default settings for prioritizing scans for follow-up review.

To reduce ambiguity and catch clear pipeline failures, the dashboard also reports ``hard'' QC checks including the fraction of non-finite voxels (\texttt{NaNInf\_Frac}), the fraction of zero-valued voxels in the full image (\texttt{Zero\_Frac}), the fraction of voxels included in the non-zero analysis region (\texttt{Coverage}), the fraction of negative voxel intensities within that region (\texttt{NegFrac}), and the clipping fraction within that region (\texttt{clip\_frac}). These checks help identify corrupted images, empty or near empty volumes, suspicious intensity scaling, and possible saturation or clipping artifacts.

\subsubsection{Interactive QC Review, Data Export, and Metadata Inspection}

\begin{figure}
    \centering
    \includegraphics[width=0.95\linewidth]{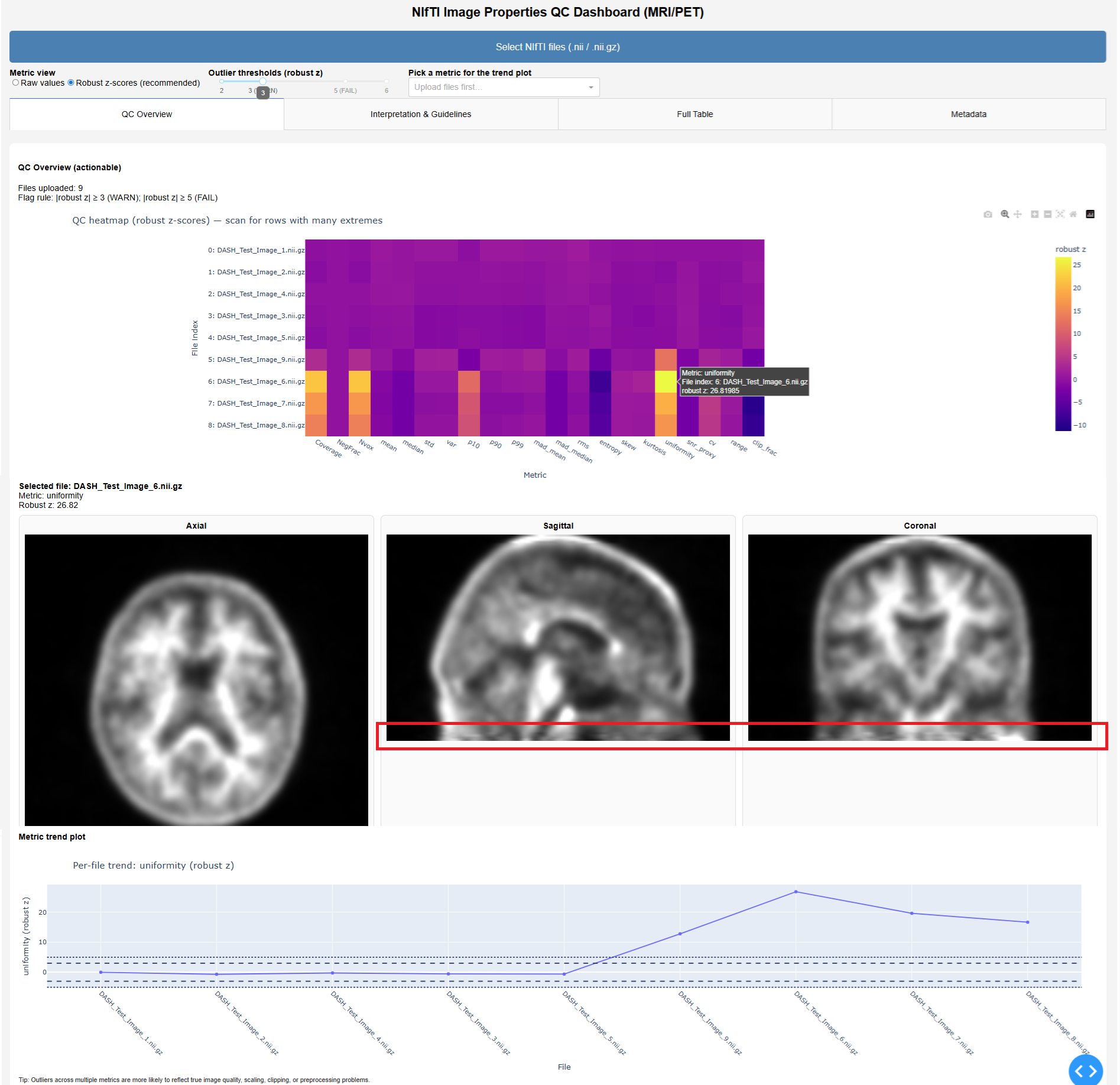}
    \caption{\textbf{QC overview of the NIfTI Image Properties dashboard.} The dashboard summarizes QC metrics for nine randomly selected PET images from the Dallas Lifespan Brain Study dataset using a robust-z heatmap and metric specific trend plot. Selecting an outlying heatmap cell links the quantitative metric to orthogonal axial, sagittal, and coronal image views, enabling rapid visual review of the flagged scan.}
    \label{fig:artifact-dashboard-1}
\end{figure}

Figure~\ref{fig:artifact-dashboard-1} shows the QC metric computation demonstration for nine randomly selected test PET images. The dashboard uses linked interactive views to support rapid review at both the batch and file levels. In the \textit{QC Overview} tab, a heatmap summarizes image property values across files, while a metric specific trend plot shows how the selected metric varies across the uploaded set under user defined raw-value or robust-z displays and outlier thresholds. Together, these views help distinguish isolated deviations from broader patterns affecting multiple scans and prioritize flagged files for review.

A key feature is the direct link between quantitative outlier detection and visual inspection: selecting a heatmap cell opens the corresponding scan in orthogonal axial, sagittal, and coronal views, enabling immediate manual assessment of whether the flagged metric is associated with a visible image quality issue. In Figure~\ref{fig:artifact-dashboard-1}, the selected scan shows an outlying \texttt{uniformity} value with subtle blurring/smoothing visible in the sagittal and coronal views.

\FloatBarrier

In addition to interactive review, the dashboard provides a structured tabular output for downstream use of QC results. As shown in Supplementary Figure 2, the \textit{Full Table} tab presents the complete per-file QC summary in a sortable and filterable format, allowing the user to inspect individual scans, compare values across files, and identify which metrics contributed to a flag.

For integration into broader workflows, the dashboard includes a \textit{Download CSV} option within the \textit{Full Table} tab (Supplementary Figure 2). This export provides the full QC table for the uploaded batch in a standard tabular format, enabling offline review, record keeping, or incorporation into external preprocessing and analysis pipelines.

And finally to improve transparency of acquisition level information, if DICOM files are available, we implemented a dedicated \textit{Metadata} tab within the dashboard for viewing file header information from imported DICOM files. As illustrated in Supplementary Figure 1, this tab provides a structured, sortable view of available metadata fields and their corresponding values across uploaded files, enabling rapid inspection of acquisition related information within the same interface used for image based QC. The tab also includes an option to export the extracted metadata in JSON format for documentation, offline review, or downstream integration with external workflows. This functionality is intended to facilitate convenient access to metadata and manual checking of acquisition consistency, rather than to provide a full harmonizability assessment based on DICOM parameters.

\subsection{Core Features}

\begin{figure}[ht]
    \centering
    \includegraphics[width=\linewidth]{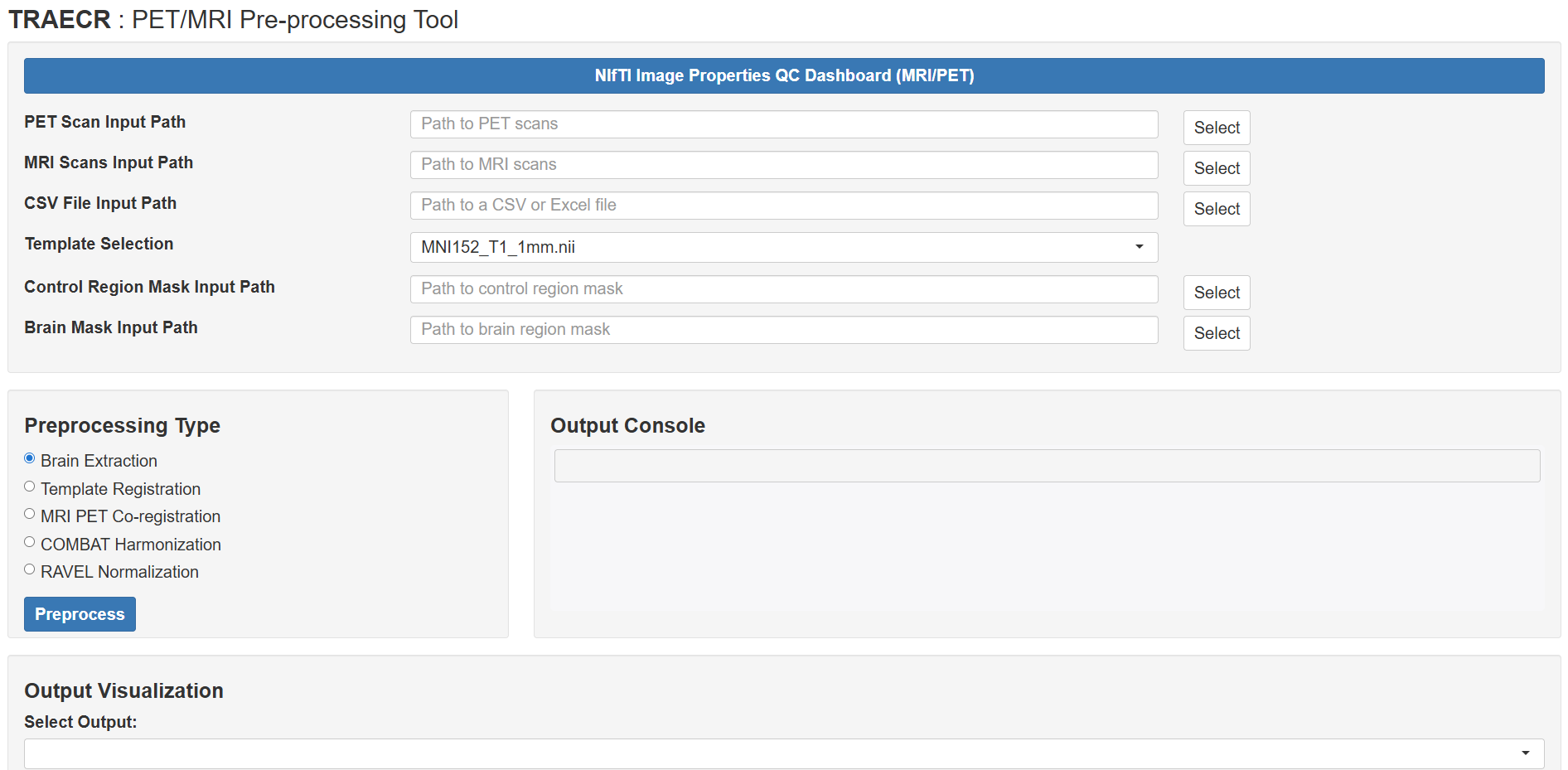}
    \caption{\textbf{Default landing page of TRAECR}
    The header bar (dark blue) links to the artifact viewer module.  
    \textit{Top block}: five file chooser rows accept PET volumes, MRI volumes, a CSV file path, a template (drop-down pre-populated with standard MNI brains), and control/brain mask selections for RAVEL normalization.  
    \textit{Middle left}: a radio panel lets the user pick one of five pre-processing pipelines (brain extraction through RAVEL normalization); pressing \textit{Preprocess} launches the selected workflow.  
    \textit{Middle right}: the \textit{Output Console} streams real-time status messages and error reports.  
    \textit{Bottom block}: once processing is complete, outputs appear in a drop-down menu; the slice slider and navigation buttons (previous/next, play/pause) enable frame-by-frame or cine review of any intermediate or final image.}
    \label{fig:default-UIPage}
\end{figure}

TRAECR is designed to streamline and enhance the processing of PET and MRI neuroimaging data by integrating multiple pre-processing functionalities into a user-friendly interface. The tool’s home page, with all core features, is shown in Figure \ref{fig:default-UIPage}. These features include:

\begin{itemize}

\item \textbf{Brain Extraction}: Perform brain extraction on MRI scans to isolate brain tissue, removing non-brain elements such as the skull, scalp, and background voxels. This step is essential for accurate subsequent analyses, such as registration and parcellation (segmentation of structural brain regions).

\item \textbf{MNI Template Registration}: Register brain extracted images to the MNI template space using a selectable template from a drop-down menu. This facilitates standardized comparisons across subjects and studies by aligning images to a common anatomical space.

\item \textbf{MRI-PET Co-registration}: Perform co-registration of PET images with corresponding MRI scans using a CSV input file containing the paths to MRI and PET image pairs. This enables precise alignment of functional (PET) and anatomical (MRI) data, allowing multimodal analyses as well as enabling population-level, voxel-level analyses of PET images.

\item \textbf{COMBAT Harmonization}: Implementation of the COMBAT harmonization method to adjust for batch effects and scanner induced variability. Users provide a CSV file containing batch or scanner details for each input image to facilitate this process, improving data comparability across different scanners or study sites.

\item \textbf{RAVEL Normalization}: Incorporate RAVEL normalization to reduce scanner related non-biological differences among images. Users provide the location of the control region masks and brain masks for this process, ensuring that genuine biological variations are retained in the data.

\item \textbf{Output Message Panel}: An output message panel where results and messages are communicated to the user. This feature provides feedback on the processing status, alerts users to any issues encountered, and enhances the overall user experience by keeping them informed throughout the workflow.

\item \textbf{Report Generation}: Automated generation of a self contained, downloadable HTML report for each processing step, capturing (i) the selected workflow/step name and timestamp, (ii) complete input/output file inventories (paths, sizes, modified times), (iii) key parameters used for the run, and (iv) practical QC summaries—including header/geometry checks, intensity sanity checks (e.g., NaNs/Infs/negatives), empty-output detection, similarity metrics (normalized cross correlation and mutual information when images are on the same grid), and basic affine registration checks (translation/scale/determinant). The report also embeds quick-look figures (mid-slice snapshots and overlays/contours when applicable), includes the full processing log, and records session information to support transparency and reproducibility.

\item \textbf{Visualization and Data Analysis}: An interactive visualization feature, allowing users to view input images, intermediate pre-processing steps, and final preprocessed images side-by-side. When multiple cases are processed in a batch, the output viewer drop-down lists each processed case, allowing users to switch between outputs and review results case-by-case as shown in Supplementary Figure 4. Additionally, users can use play/pause controls to automatically scroll through slices for rapid visual QC, or pause at any slice to examine local artifacts and registration quality in detail.

\end{itemize}

\FloatBarrier

\section{Algorithms and Techniques}\label{s:Algorithms}

In this section, we concisely describe the core preprocessing steps available in our tool using a publicly available dataset from the Dallas Lifespan Brain Study \citep{park2025dallas} as an example. These steps include brain extraction, MNI template registration, MRI to PET co-registration, and post-acquisition harmonization and normalization using COMBAT and RAVEL.

\subsection{Brain Extraction}

\begin{figure}
    \centering
    \includegraphics[width=0.9\linewidth]{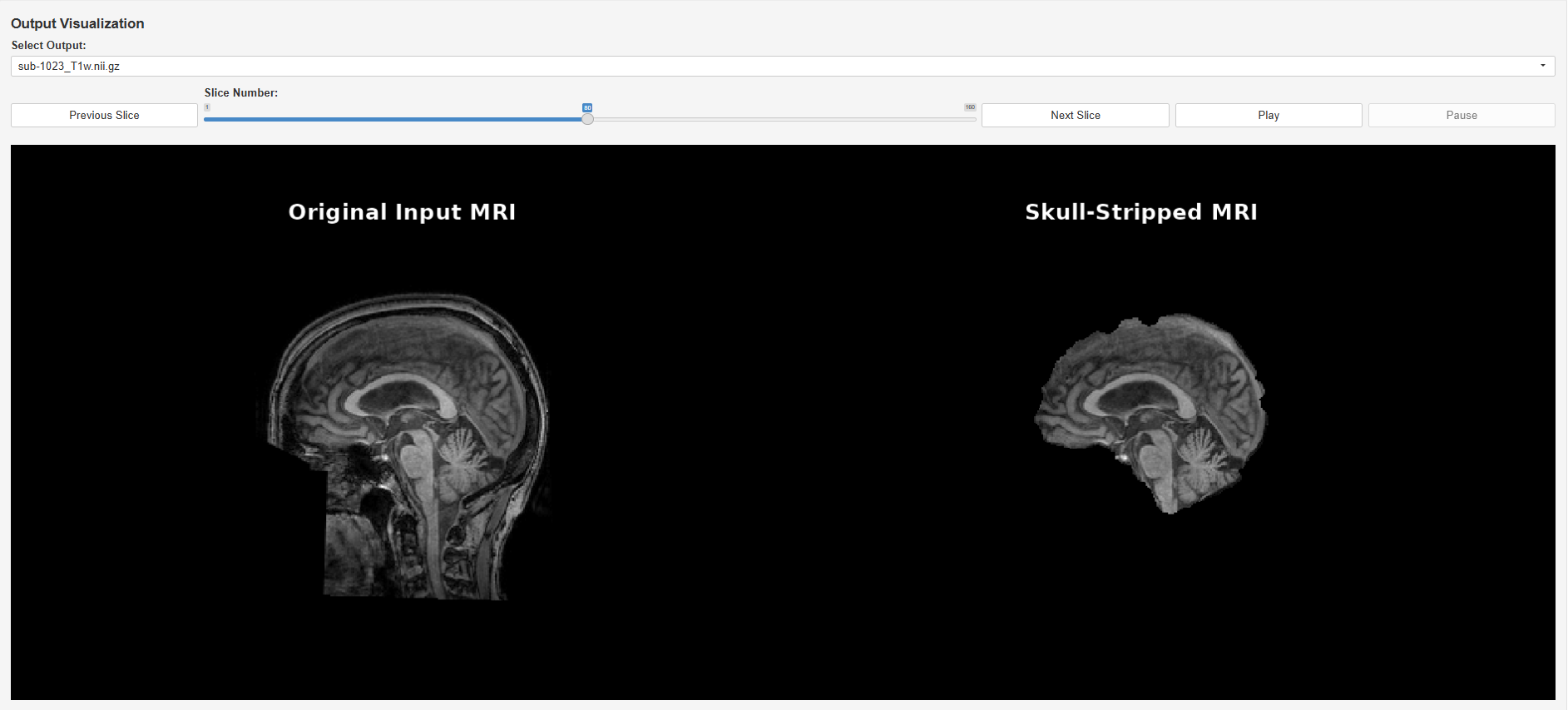}
    \caption{\textbf{Interactive verification of the Brain-Extraction module.}
    The drop-down menu (top) lets the user choose any processed file.  
    The slice slider and navigation buttons enable frame-by-frame or cine browsing.  
    \textit{Left panel}: original T1-weighted MRI slice.  
    \textit{Right panel}: skull-stripped output generated by the brain extraction
    workflow.  
    Side-by-side display allows immediate visual confirmation that non-brain
    tissue has been removed before downstream processing proceeds.}
    \label{fig:brain-extraction}
\end{figure}

Accurate brain extraction is a crucial pre-processing step for most neuroimaging tasks. In our software tool, we incorporate a robust brain extraction feature that enables the isolation of brain tissue from MRI scans by removing non-brain elements like the skull and scalp. When brain extraction is selected, the tool accepts one or more MRI file paths specified by the user. The paths are normalized to ensure consistent file handling across different operating systems. For each MRI file, the tool logs the filename, skull-strips the image with \texttt{fslbet\_robust(remover = 'double\_remove\_neck')}, creates an output directory named after the file (without file extension), saves the processed volume, and reports completion in the console. Figure \ref{fig:brain-extraction} displays the visualization panel of the brain extraction feature. The panel shows the non-skull-stripped brain image on the left and the skull-stripped brain image on the right, enabling visual comparison. Additionally, the interface features a navigation bar and an image selection dropdown menu, enabling users to easily browse and select images for visualization. 

\FloatBarrier

\subsection{MNI Template Registration}

Registering MRI scans to a standardized anatomical space, such as the MNI template, is essential for comparative analyses across subjects and studies. Our tool provides a template registration feature to facilitate this alignment process. When MNI registration is selected, the software accepts one or more MRI file paths specified by the user. The tool then logs the inputs, prepares the chosen MNI template, and makes an output folder; both MRI and template are skull-stripped, then the MRI is registered to the template with FLIRT using an affine transformation, which is computationally efficient, robust across heterogeneous acquisitions, and provides sufficient global alignment for downstream QC visualization and ROI level analyses. The resulting transformation matrix and registered MRI are saved, and the output console confirms its completion.

\begin{figure}
    \centering
    \includegraphics[width=0.9\linewidth]{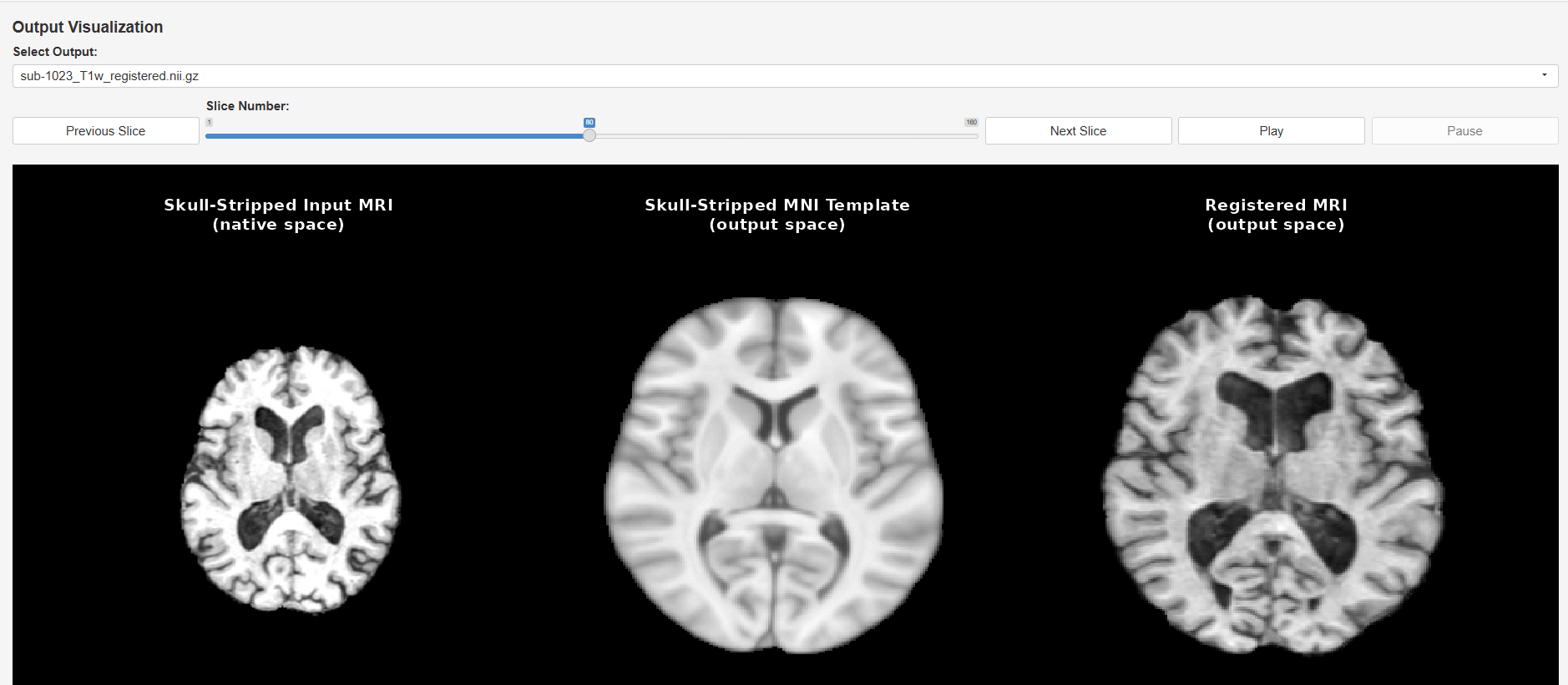}
    \caption{\textbf{Interactive review of MNI-space registration module}
    The drop‐down selector (top) shows the registered volume.  
    A slice slider and navigation buttons provide frame-by-frame or cine browsing.  
    \textit{Left}: skull-stripped input MRI, shown at an approximately matched slice.  
    \textit{Center}: skull-stripped MNI template chosen from the
    template menu.  
    \textit{Right}: input MRI after affine alignment to MNI space,
    allowing immediate visual confirmation that gross anatomical structures
    coincide with the template before downstream analysis.}
    \label{fig:mni-registration}
\end{figure}

Figure \ref{fig:mni-registration} displays the visualization panel of the MNI template registration feature. The panel shows the skull-stripped brain image (left), skull-stripped template image (middle) and template registered MRI (right), allowing for visual comparison of the steps performed to generate the required results.

An optional atlas-based parcellation feature can be enabled for this module to support ROI-level summaries in subject space. When selected, TRAECR automatically selects the labeled atlas file (i.e., AAL) made available in the tools \texttt{templates/} directory and propagates the atlas labels into the subject’s MRI space using the inverse of the MRI$\rightarrow$MNI affine transformation estimated during registration. Specifically, the tool computes the MNI$\rightarrow$MRI transform and resamples the atlas into the skull-stripped subject MRI grid using nearest-neighbor interpolation to preserve discrete labels. The resampled label map, per ROI binary masks, and a CSV of mean ROI intensities are saved to the subject-specific output folder. This enables downstream regional analyses and transparent QC of ROI definitions in native MRI space.

 \FloatBarrier

\subsection{MRI-PET Co-registration}

Integrating multimodal neuroimaging data, such as MRI and PET scans, requires precise alignment to ensure accurate overlay of anatomical and functional information. Our tool includes an MRI-PET co-registration feature that aligns PET images to MRI images. When MRI-PET co-registration is selected, the software requires a CSV or an Excel file specifying pairs of MRI and PET image paths. This file must include two essential columns:

\begin{itemize} \item \texttt{MRI\_InputPath}: Full paths to the MRI images. \item \texttt{PET\_InputPath}: Full paths to the corresponding PET images. \end{itemize}

TRAECR normalizes the file paths for consistent access and verifies the existence of all specified images, reporting any missing files to the user. For each MRI–PET pair, the pipeline first loads the MRI and PET files, skull–strips the MRI to obtain a binary brain mask, and registers the native PET to the native MRI using FLIRT with an affine transformation. The pipeline then registers the skull–stripped MRI mask volume to the skull–stripped template to obtain the MRI\(\rightarrow\)template transformation matrix (FLIRT, affine transformation; \texttt{mri.mat}). Next, the tool applies the MRI–space brain mask obtained during the brain extraction step to the PET, and finally applies the MRI\(\rightarrow\)template transformation matrix to the masked PET, yielding a skull–stripped PET in template space. All outputs are saved with final status messages logged in the output console.

For MRI--PET co-registration, TRAECR additionally supports atlas based parcellation in the subject’s \emph{native PET space} when the parcellation option is enabled. The pipeline composes the estimated transforms to map atlas labels from MNI space directly into PET space: the MRI$\rightarrow$MNI and PET$\rightarrow$MRI affine transformation matrices are inverted to obtain MNI$\rightarrow$MRI and MRI$\rightarrow$PET transformations, respectively, and these are concatenated to yield an MNI$\rightarrow$PET transform. The atlas is then resampled into the original PET grid using nearest-neighbor interpolation, and ROI-wise mean values are computed on the PET image in that same space. The tool writes the PET space label map, per-ROI masks, and a CSV table of ROI means to the output folder. This step ensures that regional PET intensity summaries and parcellation visualizations are generated in a consistent, reproducible manner.

\begin{figure}
    \centering
    \includegraphics[width=0.9\linewidth]{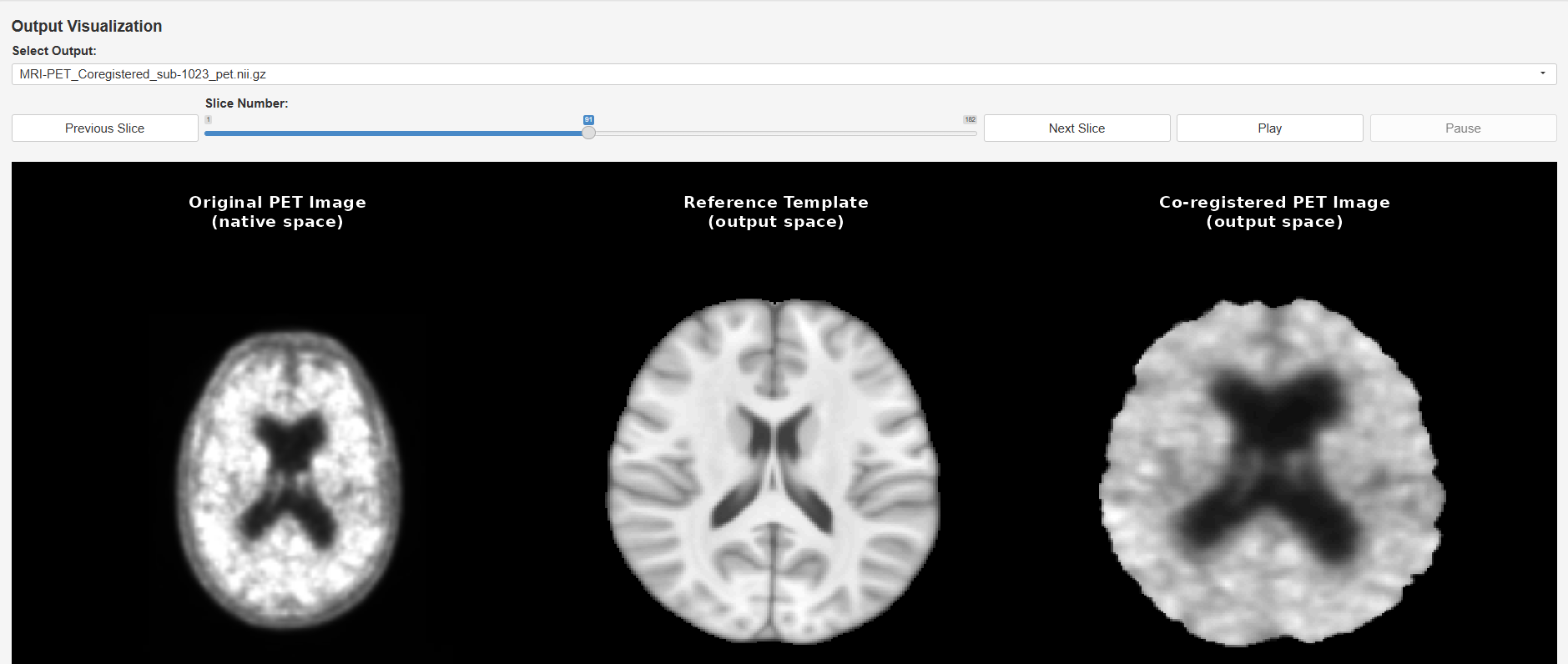}
    \caption{\textbf{Interactive review of MRI–PET co-registration module}
    The drop-down menu (top) displays the co-registered output.  
    The slice slider plus navigation buttons allow step-wise or cine browsing through the volume. 
    \textit{Left}: original PET image in scanner space, shown at an approximately matched slice.
    \textit{Center}: skull-stripped MNI template chosen from the
    template menu.
    \textit{Right}: PET image after affine alignment to the subject’s
    MRI and resampling into MRI voxel space ready for voxel-wise multimodal
    analyses.}
    \label{fig:mri-pet-coreg}
\end{figure} \FloatBarrier

Figure~\ref{fig:mri-pet-coreg} illustrates the output visualization panel for the MRI–PET co-registration module.  The left panel shows the original PET slice, whereas the right panel shows the same slice after affine alignment and resampling into the subject’s MRI space.  
Users can move through the volume with the slice slider or cine controls above the images to verify the accuracy of the registration across all slices.

\subsubsection{Validation Using The Dallas Lifespan Brain Study Wave 3}

\begin{figure}[ht]
\centering
\includegraphics[width=0.95\linewidth]{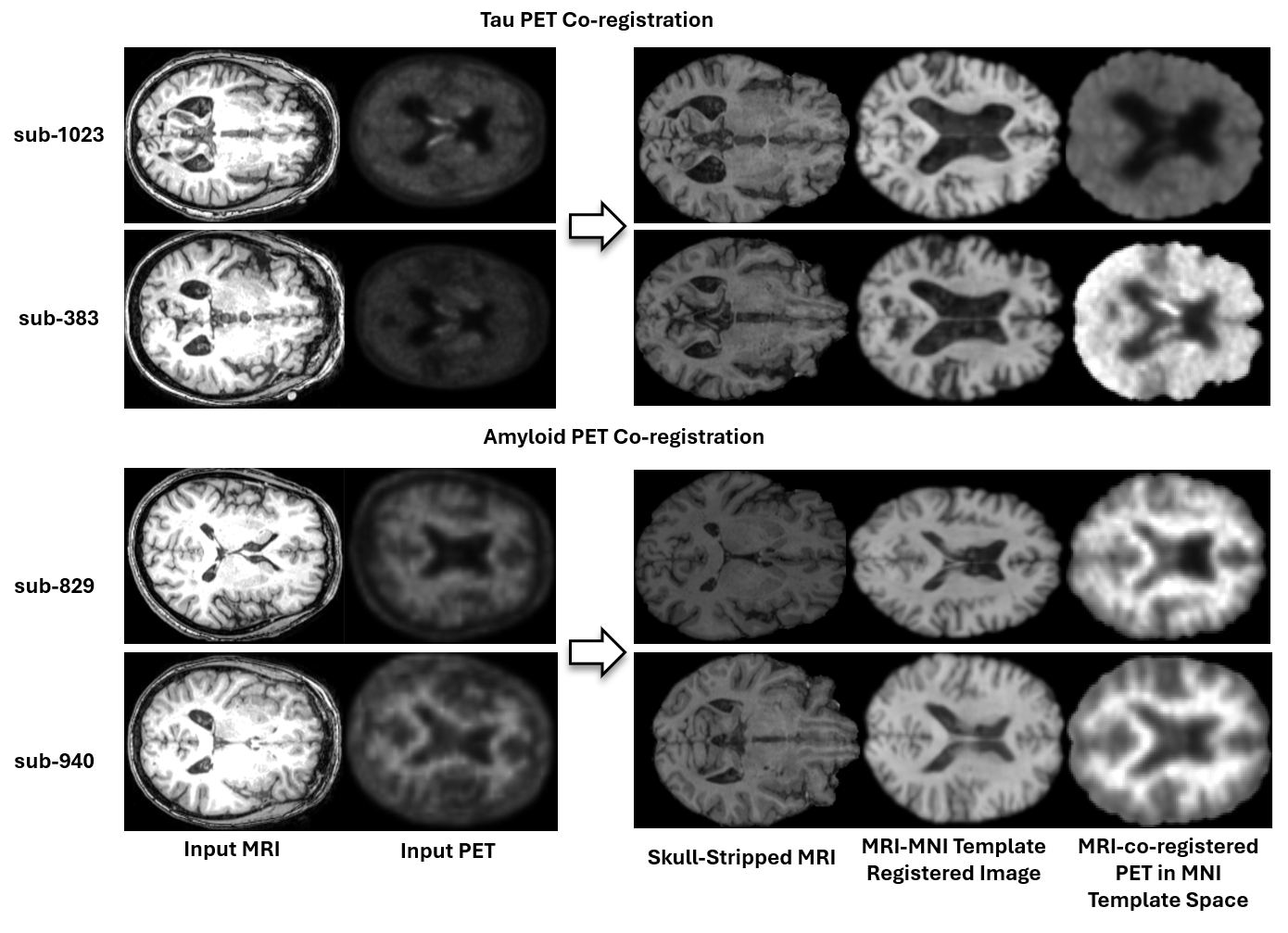}
\caption{\textbf{Representative MRI--PET preprocessing outputs from The Dallas Lifespan Brain Study Wave 3 validation dataset.}
Axial slices are shown for representative tau PET cases and amyloid PET cases processed using the TRAECR workflow. For each subject, the left panels show the input structural MRI and input PET image, while the right panels show the skull-stripped MRI, MRI registered to the MNI template, and the MRI-coregistered PET image transformed to MNI template space.}
\label{fig:dlbs-external-validation-combined}
\end{figure}

To evaluate the image QC dashboard and core MRI--PET pre-processing components of TRAECR, we applied the workflow to Wave 3 MRI and PET images from The Dallas Lifespan Brain Study \citep{park2025dallas} dataset. The validation focused on the QC dashboard, brain extraction, MRI to template registration, and MRI--PET co-registration steps. Although the dataset publication reports MRIQC-based assessment of T1-weighted MRI quality and manual quality checking/editing of FreeSurfer derived structural outputs, these procedures do not directly assess PET image quality. Therefore, prior to pre-processing, image level QC was performed independently using the TRAECR QC dashboard for the Wave 3 amyloid PET and tau PET images. Here, no scans showed visual brain image quality failures that required exclusion; therefore, all reviewed scans were retained for pre-processing. Additional details and representative QC results are provided in Section 2 (Figures 5-10) of the supplementary material. The validation dataset consisted of 134 subjects and 138 structural MRI scans, with four subjects having two MRI scans. The PET data included 76 amyloid PET scans and 121 tau PET scans. In total, 203 MRI--PET pairings were pre-processed through the TRAECR workflow, including MRI--amyloid PET and MRI--tau PET pairs. Representative examples of the validation pre-processing workflow for amyloid PET and tau PET are shown in Figure~\ref{fig:dlbs-external-validation-combined}. Harmonization was not evaluated in this validation, as the Dallas Lifespan Brain Study analysis used Wave 3 data from a single study with pre-processing performed separately for each tracer. Thus, there was no defined cross-study, cross-wave, or multi-batch harmonization target for COMBAT or RAVEL.

The evaluated pre-processing workflow for all 203 MRI--PET pairings was completed in approximately five hours on a Dell Precision 3660 workstation running Windows 11 Enterprise, with a 13th Gen Intel Core i9-13900 processor, 24 CPU cores/32 logical processors, and 64 GB RAM. This corresponds to an approximate processing time of less than 1.5 minutes per MRI--PET pair.

\FloatBarrier

\begin{figure}[ht]
\centering
\includegraphics[width=0.95\linewidth]{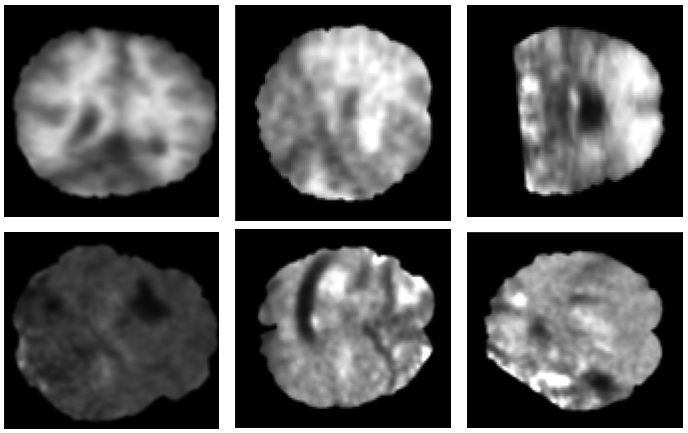}
\caption{\textbf{Representative MRI--PET co-registration failure examples from the validation dataset.} Axial mid-slice outputs are shown for cases in which the affine registration converged to suboptimal anatomical alignment despite use of the same standardized TRAECR registration workflow. The top row shows failed amyloid PET examples, and the bottom row shows failed tau PET examples. These examples demonstrate the need for post-registration visual quality control to identify failed or unreliable co-registration outputs before downstream analysis.}
\label{fig:dlbs-registration-failure}
\end{figure}

Although the TRAECR workflow was successfully executed across the validation dataset, post-registration QC remained necessary because affine registration can occasionally converge to suboptimal solutions for individual MRI--PET pairs. Representative failed co-registration examples are shown in Figure~\ref{fig:dlbs-registration-failure}. Potential contributing factors may include poor initial PET--MRI alignment, header or orientation inconsistencies, field-of-view differences, low PET anatomical contrast, motion or reconstruction artifacts, and extracranial signal influencing the affine registration cost function. Failed co-registration was observed in 15 of 76 amyloid PET registrations (19.7\%) and 33 of 121 tau PET registrations (27.3\%).

To evaluate whether the TRAECR QC dashboard could help identify problematic post-registration outputs across all Wave 3 output files, QC metrics were also computed for the registered PET images. Failed registration cases showed abnormal robust-z values for the coverage metric, whereas visually successful registrations generally remained within the expected range. Thus, the dashboard could potentially be used to prioritize outputs requiring closer review, while visual inspection remains useful as a final confirmation step. Representative dashboard examples of failed and successful post-registration outputs are provided in Section 2.1 (Figures 11-14) of the supplementary material. These results highlight the importance of post-registration QC, either through the TRAECR dashboard or visual inspection, before downstream analysis.

\subsection{COMBAT Harmonization}

In neuroimaging, particularly in studies that span multiple locations or occur over long periods, the presence of batch effects can significantly challenge the analysis and interpretation of the data. These batch effects often arise due to variations in scanner hardware, software, or other technical aspects of image acquisition across different sites. To address this, our tool employs COMBAT harmonization \citep{fortin2018harmonization} which is adept at modeling and correcting for these sources of technical variability.

In addition to image acquisition variables, we recommend providing biological covariates that should be preserved during harmonization, when available, such as \texttt{Diagnosis}, \texttt{Age}, \texttt{Sex}, and \texttt{Visit}. As another example, in PET studies that include data collected with more than one tracer, tracer related differences should be considered explicitly during harmonization, since they may otherwise be conflated with \texttt{Batch}; depending on the study design and analysis goal, this may motivate harmonizing separately by tracer or evaluating whether \texttt{Tracer} should be modeled in the design matrix \citep{yang2024evaluation}. The provided covariates are included in the COMBAT design matrix so that scanner/site effects are reduced while variation attributable to specified biological factors is retained. If a biological factor of interest is unevenly distributed across \texttt{Batch} and omitted from the model, COMBAT may partially attenuate true biological signal by attributing part of that variation to the batch. Accordingly, covariate distributions across \texttt{Batch} should be reviewed, and variables whose effects should be retained should be included in the COMBAT design matrix. For example, in a multi-site case control study where diagnosis differs by site, including \texttt{Diagnosis} in the COMBAT model may help reduce the risk of removing true disease related effects associated with \texttt{Batch}.

In TRAECR, when COMBAT harmonization is selected, the tool requires a CSV or an Excel file containing covariate information for each PET image. This file must include the below essential columns:

\begin{itemize} \item \texttt{Filename}: Full paths to the PET images to be harmonized. \item \texttt{Batch}: Identifiers indicating the batch or scanner site associated with each image. \end{itemize}

TRAECR reads the covariate file and normalizes the file paths to ensure consistent access across different operating systems. It verifies the existence of all specified PET image files, reporting any missing files to the user.

For each image, the tool loads the PET volume using \texttt{readNIfTI} from the \texttt{oro.nifti} R package, vectorizes the 3D voxel intensities, and assembles them into a data matrix $\mathbf{X}\in\mathbb{R}^{p\times n}$ (voxels $\times$ subjects). Voxels with zero variance across subjects are discarded to form $\mathbf{X}'$, avoiding computational issues (i.e. resulting in shorter computation time) and irrelevance to batch effect estimation. The \texttt{neuroCombat} function (from the \texttt{neuroCombat} R package) is then applied to $\mathbf{X}'$ using batch labels, removing technical variability while preserving biological signal and yielding $\mathbf{X}'_{\mathrm{combat}}$ with the same dimensions as $\mathbf{X}'$. Constant (removed) features are reinserted at their original values to reconstruct $\mathbf{X}_{\mathrm{combat}}$. These voxels which are almost exclusively background zeros outside the analysis mask are restored unchanged only to preserve image geometry; because they are identical across subjects and excluded from statistical analyses, they neither introduce outliers nor bias estimates. The result is reshaped back to the original 3D geometry for each PET image with the NIfTI header metadata retained. Harmonized images are written to a designated output folder with filenames denoting their processed status, and the console continuously reports progress and any encountered issues.

\begin{figure}[ht]
    \centering
    \includegraphics[width=\linewidth]{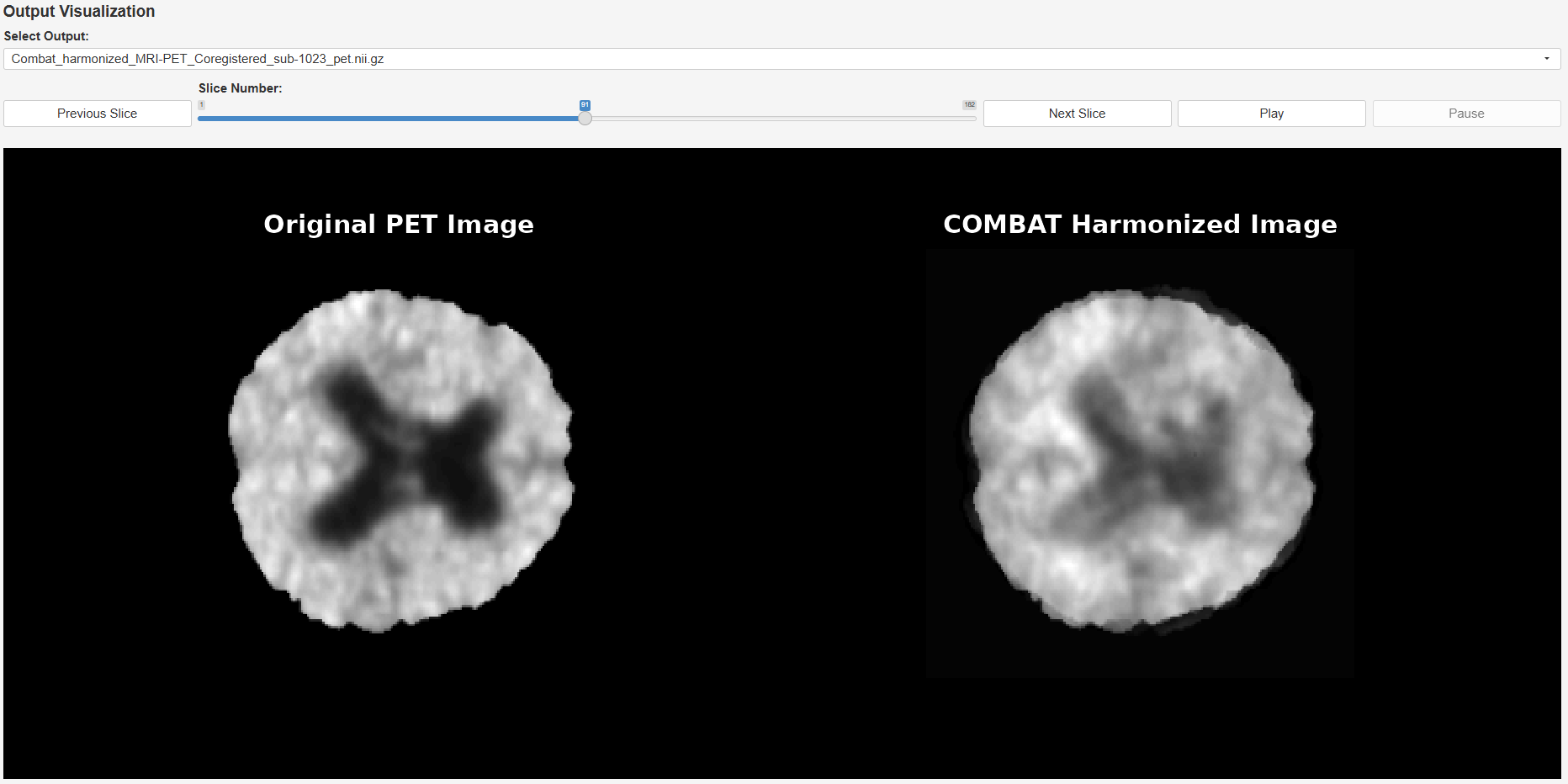}
    \caption{\textbf{Interactive review of COMBAT harmonization module}
    The drop-down menu (top) displays the harmonized output; a slice slider and
    navigation buttons allow slice by slice inspection.  
    \textit{Left}: original PET slice acquired on one scanner.  
    \textit{Right}: same slice after COMBAT has removed scanner specific bias
    by applying location and scale parameter adjustments derived from the
    study’s batch design matrix.}
    \label{fig:combat-harmonisation}
\end{figure}

Figure~\ref{fig:combat-harmonisation} presents the COMBAT harmonization viewer. On the left is the co-registered PET scan exactly as acquired; on the right is the same volume after COMBAT has removed scanner specific location and scale biases. Because COMBAT re-centres and re-scales voxel intensities rather than altering anatomy, structural details remain visually unchanged yet the underlying intensity distribution is now standardized across scanners, enabling unbiased group analyses.

\FloatBarrier

\subsection{RAVEL Normalization}

Scanner induced anomalies or scanner drift over time can introduce technical variability that distorts true biological signals. To address this, our tool incorporates RAVEL normalization \citep{fortin2016removing}. This method is designed to remove unwanted variability from PET images while preserving biological signal.

When RAVEL normalization is selected, the tool requires the following inputs:

\begin{itemize} \item \texttt{PET Images}: Full paths to the PET images to be normalized. \item \texttt{Brain Mask}: A binary mask defining brain tissue voxels. \item \texttt{Control Mask}: A binary mask specifying control regions unaffected by pathology, used to model technical variability. \end{itemize} 

RAVEL assumes that intensities within the \texttt{Control Mask} primarily reflect unwanted technical variation (e.g., scanner drift or global scaling) and are minimally affected by the biological process under study. Therefore, the control region should be selected to be pathology insensitive for the specific tracer/cohort (e.g., an established reference region or another stable region appropriate to the application). If the control region is affected by disease or differs systematically across groups, RAVEL may attenuate biological effects. Accordingly, the documentation emphasizes careful selection of a pathology insensitive control region for the specific tracer/cohort and recommends verifying that control region uptake is stable across subjects and groups before applying RAVEL.
For example, if the control region is impacted by disease related uptake changes, RAVEL may reduce true group differences along with scanner related variability.

TRAECR reads all three inputs, normalizes their file paths, confirms file existence, and checks that both masks match the PET images in size and orientation. For each input image, the tool loads the PET volume and its brain/control masks with \texttt{readNIfTI}, verifies that the masks match the PET dimensions, and applies the RAVEL algorithm to adjust voxel intensities within the brain mask using control-region intensities to remove technical variability. The normalized values are written back to brain voxels while voxels outside the brain mask are set to zero; the resulting images are saved in a designated output directory with filenames indicating their processed status, and console messages report progress and any issues.

\begin{figure}[ht]
    \centering
    \includegraphics[width=\linewidth]{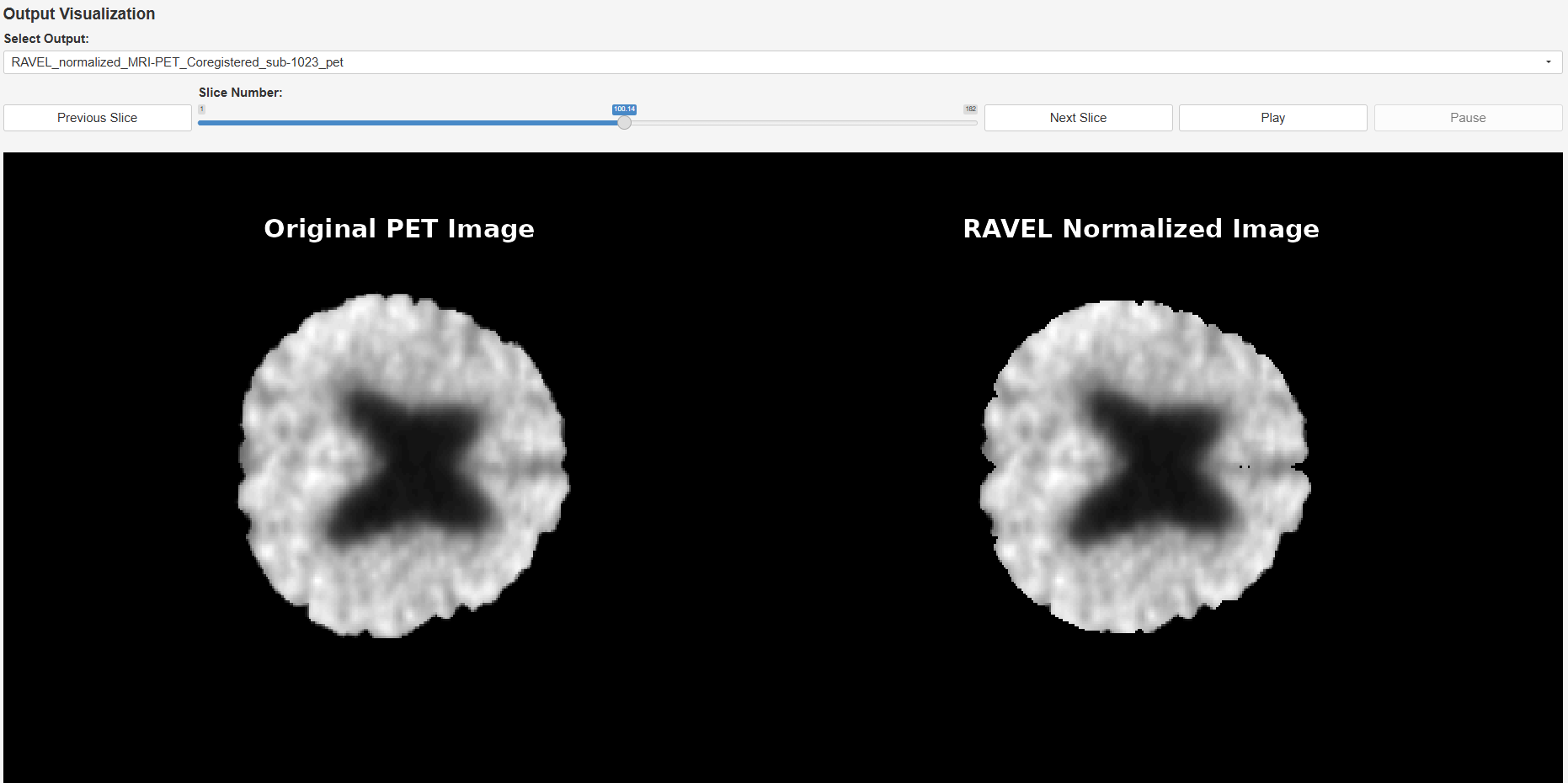}
    \caption{\textbf{Interactive review of RAVEL normalization module}
    The drop-down selector (top) lists the normalized output path; the slice slider and
    navigation buttons allow slice by slice inspection.  
    \textit{Left}: original PET slice after co-registration.  
    \textit{Right}: PET slice after RAVEL has regressed out signal variation
    linked to the user-specified control region mask, thereby reducing
    scanner-dependent intensity drift while preserving biological contrast.  
    Structural appearance remains largely unchanged because the
    correction operates on voxel wise residuals rather than on anatomical
    geometry; the improvement becomes evident when comparing intensity
    distributions across subjects.}
    \label{fig:ravel-normalisation}
\end{figure}

Figure~\ref{fig:ravel-normalisation} shows the RAVEL-normalization viewer. The left panel contains the co-registered PET slice exactly as input, while the right panel displays the same slice after RAVEL has regressed out scanner related variation using the designated control region. Because RAVEL adjusts voxel intensities without altering anatomy, the two images appear nearly identical yet the right-hand volume now has non-biological intensity drift removed, making cross-subject comparisons more reliable.

 \FloatBarrier

\section{Tool Setup and Deployment Procedure}\label{s:toolsetup}

To streamline deployment and ensure reproducible execution across heterogeneous user machines, TRAECR is distributed with a standardized Linux runtime environment. On Windows systems, we use Windows Subsystem for Linux (WSL2) with Ubuntu 20.04 LTS to provide a consistent Linux user space while preserving a native Windows workflow. TRAECR has been validated on Windows 10/11 (WSL2) and on native Ubuntu 20.04. On macOS, where WSL is unavailable, TRAECR can be executed in a Docker container that mirrors the same Ubuntu 20.04 environment; macOS support has not yet been formally tested and will be verified and documented in a subsequent release.

\subsection{Computational Environment Configuration}

Within the Ubuntu 20.04 environment, we install R (v4.2.0) and the R packages required for the pipeline and user interface, including \texttt{shiny} for the local web-based interface and neuroimaging utilities such as \texttt{extrantsr}, \texttt{neurobase}, and \texttt{oro.nifti}. TRAECR also incorporates Python (v3.8) for functionality not readily available in R. Interoperability between R and Python is handled via \texttt{reticulate}, which is configured to point to the intended Python installation to avoid path and dependency conflicts.

Users interact with TRAECR through the Shiny application running locally (via the WSL environment on Windows). All input, intermediate, and output neuroimaging files remain on the user’s machine; both raw and processed data are read from and written to local storage, and no data transfer to external servers is required for routine operation.

\subsection{Tool Deployment and Execution}

For Windows users, we provide two deployment options: (i) importing a pre-configured WSL2 Ubuntu 20.04 environment distributed as a \texttt{.tar} archive from a cloud-hosted location, or (ii) installing the same environment from scratch by following step-by-step instructions in the GitHub documentation. Once installed, TRAECR is launched from within the Ubuntu environment and accessed through a browser on the local machine.

\section{Conclusion and Future Work}\label{s:conclusions}

In this paper, we present TRAECR, an integrated, user-friendly pre-processing tool for reconstructed static PET volumes with corresponding MRI data. TRAECR consolidates key downstream preprocessing and quality-control steps, including artifact detection, brain extraction, template registration, MRI--PET co-registration, atlas-based parcellation, COMBAT harmonization, RAVEL normalization, QC visualization, and metadata review into a single reproducible workflow. By streamlining these steps, TRAECR reduces workflow complexity, minimizes potential processing errors, and supports more reliable and comparable neuroimaging data preparation for statistical modeling.

Future releases will focus on cross-platform compatibility validation, advanced artifact-detection methods, automated parameter selection, and support for additional imaging modalities, including functional MRI data. Future implementations will also explore the use of DICOM derived acquisition variables to inform harmonization workflows when such metadata are available. Planned extensions also include optional support for dynamic PET when frame-level data are available, including frame selection, frame aggregation, and evaluation of motion aware strategies. Finally, future versions will include additional registration back ends, including optional non-linear registration for applications requiring higher anatomical precision, along with comprehensive validation against established preprocessing pipelines and publicly available reference datasets.

\section{Tool Availability}
The software tool to support the methods detailed here along with all the necessary documentation are freely available on https://github.com/aambekar-brown/PET-Pre-processing-Tool. 

\section{Funding}
This work was supported by the National Institute of Aging under Grant R01AG075511. 

\section{Disclosure statement}
The authors report no competing interests.

\bibliographystyle{plainnat}
\bibliography{mybib}

\end{document}


\section{Image Property Quality Control Dashboard: Interpretation \& Guidelines}

\label{sec:image-property-qc-dashboard}

The image property quality control (QC) dashboard was designed to provide summary metrics for MRI and PET NIfTI images. For each uploaded NIfTI image, intensity based metrics were computed from the non-zero foreground region, defined as all non-zero voxels in the image. This foreground region should not be interpreted as brain tissue only. For skull-stripped images, the non-zero region may closely represent brain tissue; however, for non-skull-stripped MRI or PET images, it may also include skull, neck, extracranial tissue, scanner bed, padding, or reconstruction related artifacts.

\begin{figure}[ht]
    \centering
    \includegraphics[width=1\linewidth]{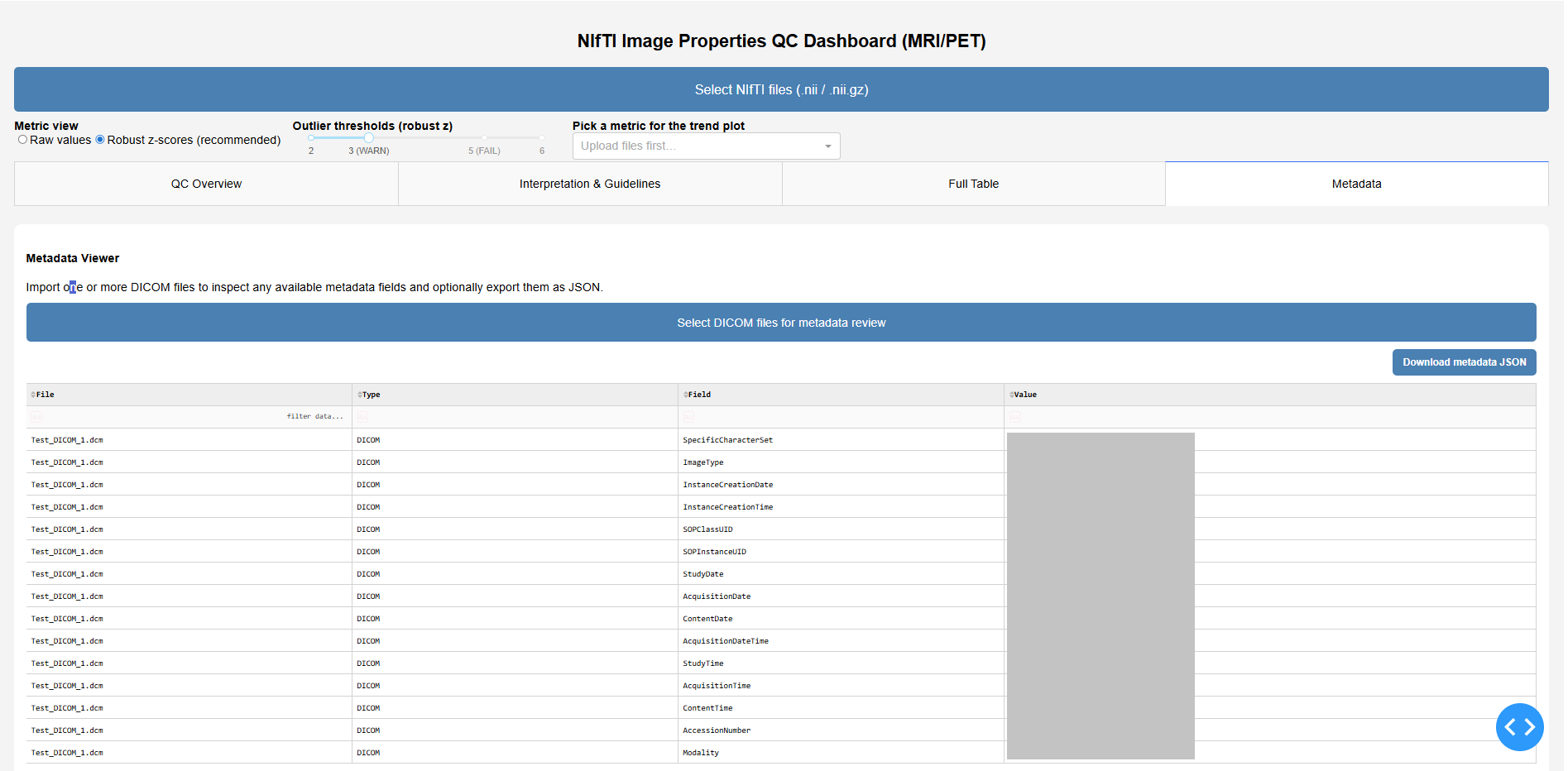}
    \caption{%
    \textbf{Metadata tab for DICOM header inspection within the dashboard.}
    This view allows users to import one or more DICOM files, inspect available metadata fields in a structured tabular format, and export the extracted header information as a JSON file. The tab is intended to support convenient review of acquisition level metadata alongside the image based QC workflow.}
    \label{fig:dicom-metadata-viewer}
\end{figure}

When DICOM files are available in addition to NIfTI images, the dashboard also provides a \textit{Metadata} tab for structured inspection of DICOM header fields and JSON export of extracted acquisition-level metadata (Figure~\ref{fig:dicom-metadata-viewer}).

\subsection{Batch Based QC Strategy}

MRI and PET intensity distributions can vary substantially across scanners, imaging sites, acquisition protocols, tracers, reconstruction settings, and preprocessing pipelines. Therefore, absolute thresholds for histogram based metrics such as entropy, skewness, kurtosis, and related intensity features are not expected to be reliable across independent studies or heterogeneous datasets. Instead, the dashboard uses a within-batch robust outlier detection strategy.

Scans were flagged according to the following rule:

\begin{itemize}
\item \textbf{WARN}: $\left|z_{\mathrm{robust}}\right| \geq 3$
\item \textbf{FAIL}: $\left|z_{\mathrm{robust}}\right| \geq 5$
\end{itemize}

This approach is intended to identify scans that differ substantially from the rest of the uploaded batch. For meaningful interpretation, the uploaded batch should consist of comparable images from the same modality, protocol, tracer, and preprocessing stage. Substantially different image types, such as T1-weighted MRI, PET, masks, different PET tracers, or different preprocessing outputs, should not be combined in the same QC batch because expected protocol differences may be incorrectly flagged as outliers.

The recommended workflow is as follows:

\begin{enumerate}
\item Upload a batch of comparable images from the same modality, protocol, tracer, and preprocessing stage.
\item Review the QC heatmap and flagged table.
\item Identify scans that are outliers across one or more metrics.
\item Prioritize scans that show extreme values across multiple metrics.
\item Visually inspect flagged scans using the orthogonal slice preview.
\end{enumerate}

\begin{figure}[ht]
    \centering
    \includegraphics[width=1\linewidth]{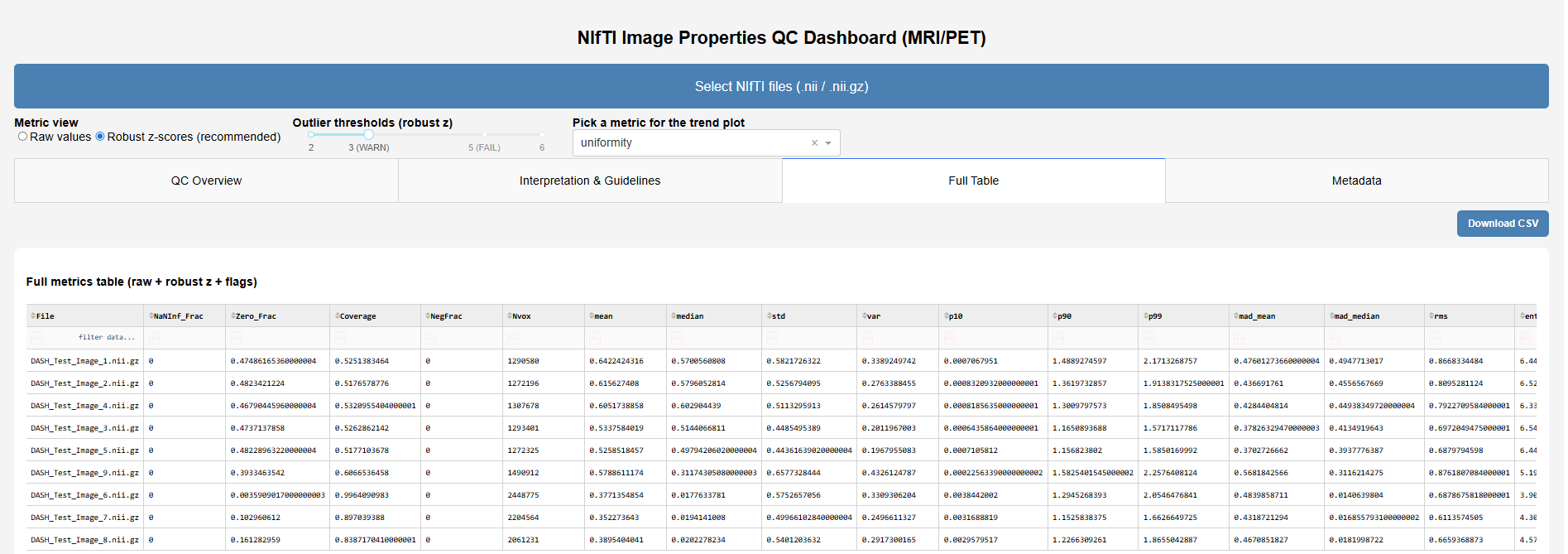}
    \caption{%
    \textbf{Full Table tab of the NIfTI Image Properties QC Dashboard.}
    This view provides a sortable and filterable table containing the complete per file QC output for the uploaded batch, including raw image-property values, corresponding robust z-scores, and QC flags. The \textit{Download CSV} button exports the full table for offline review, record keeping, or downstream analysis.}
    \label{fig:artifact-dashboard-5}
\end{figure}

The computed metrics should be interpreted as batch-level QC indicators. The complete per-file QC output, including raw metric values, robust z-scores, and QC flags, is available in the \textit{Full Table} tab, which supports sorting, filtering, and CSV export for offline review or downstream analysis (Figure~\ref{fig:artifact-dashboard-5}). They can identify scans that differ from the rest of the uploaded batch, but they do not by themselves prove the presence of a specific artifact. Final QC decisions should include visual inspection.

\subsection{Interpretation of Histogram-Based Metrics}

These metrics are computed from the intensity histogram of the non-zero foreground region.

\subsubsection{Entropy}

Entropy reflects how spread out the intensity histogram is across intensity bins. Higher entropy can indicate a broader intensity distribution, but it is not specific to one artifact type.

Entropy may be elevated when image noise is increased, when motion, ghosting, or blurring alters the intensity distribution, when intensity scaling or normalization is inconsistent, or when the image contains a mixture of tissues or non-brain foreground structures. Entropy may be very low when the image is near-empty or mostly zero, when heavy thresholding has removed much of the signal, when the image is saturated or heavily clipped, or when the uploaded file is actually a mask, label map, or mostly uniform image.

\subsubsection{Skewness}

Skewness measures asymmetry of the intensity histogram. Positive skewness, corresponding to a long right tail, may indicate hot voxels, intensity spikes, a small number of very high-intensity voxels, mis-scaling, unusual intensity normalization, or partial field-of-view effects where only high-intensity structures remain.

Negative skewness may indicate unexpected negative values or intensity shifting. However, negative values may also be expected in certain processed images, including z-scored images, harmonized images, residual images, or statistical maps. Therefore, interpretation of skewness depends strongly on the image type and preprocessing stage.

\subsubsection{Kurtosis}

Kurtosis reflects the heaviness of the histogram tails and the presence of extreme values. High kurtosis often indicates a heavy-tailed intensity distribution, extreme outlier voxels, intensity spikes, or a distribution with many typical voxels and a small number of extreme values. Very low kurtosis can occur when the intensity distribution is unusually flat, when the image has been heavily normalized, smoothed, or transformed, or when the foreground region is overly uniform.

Clipping should be assessed directly using the clipping fraction metric rather than inferred from kurtosis alone.

\subsection{Hard QC Checks}

In addition to robust outlier detection, the dashboard reports several practical QC checks that can indicate possible processing failures or unexpected image characteristics:

\begin{itemize}
\item \textbf{\texttt{NaNInf\textunderscore Frac}}: Fraction of voxels with non-finite values. Any non-finite values are usually indicative of a pipeline failure or invalid image output.
\item \textbf{\texttt{Zero\textunderscore Frac} / \texttt{WHOLE\textunderscore Coverage}}: Extreme zero fraction or abnormal foreground coverage may suggest cropping, padding, incorrect field of view, wrong orientation, or an incorrect file type.
\item \textbf{\texttt{clip\textunderscore frac}}: A large fraction of voxels at the maximum intensity value suggests possible saturation, clipping, or rescale problems.
\item \textbf{\texttt{WHOLE\textunderscore NegFrac}}: Negative values may be suspicious for PET-like data, although they can be expected in some processed MRI, normalized, residual, harmonized, or statistical images.
\item \textbf{\texttt{WHOLE\textunderscore Nvox}}: Very low foreground voxel count may indicate an empty image, failed preprocessing, severe cropping, or an uploaded mask or label file instead of an intensity image.
\end{itemize}

\begin{figure}[ht]
    \centering
    \includegraphics[width=1\linewidth]{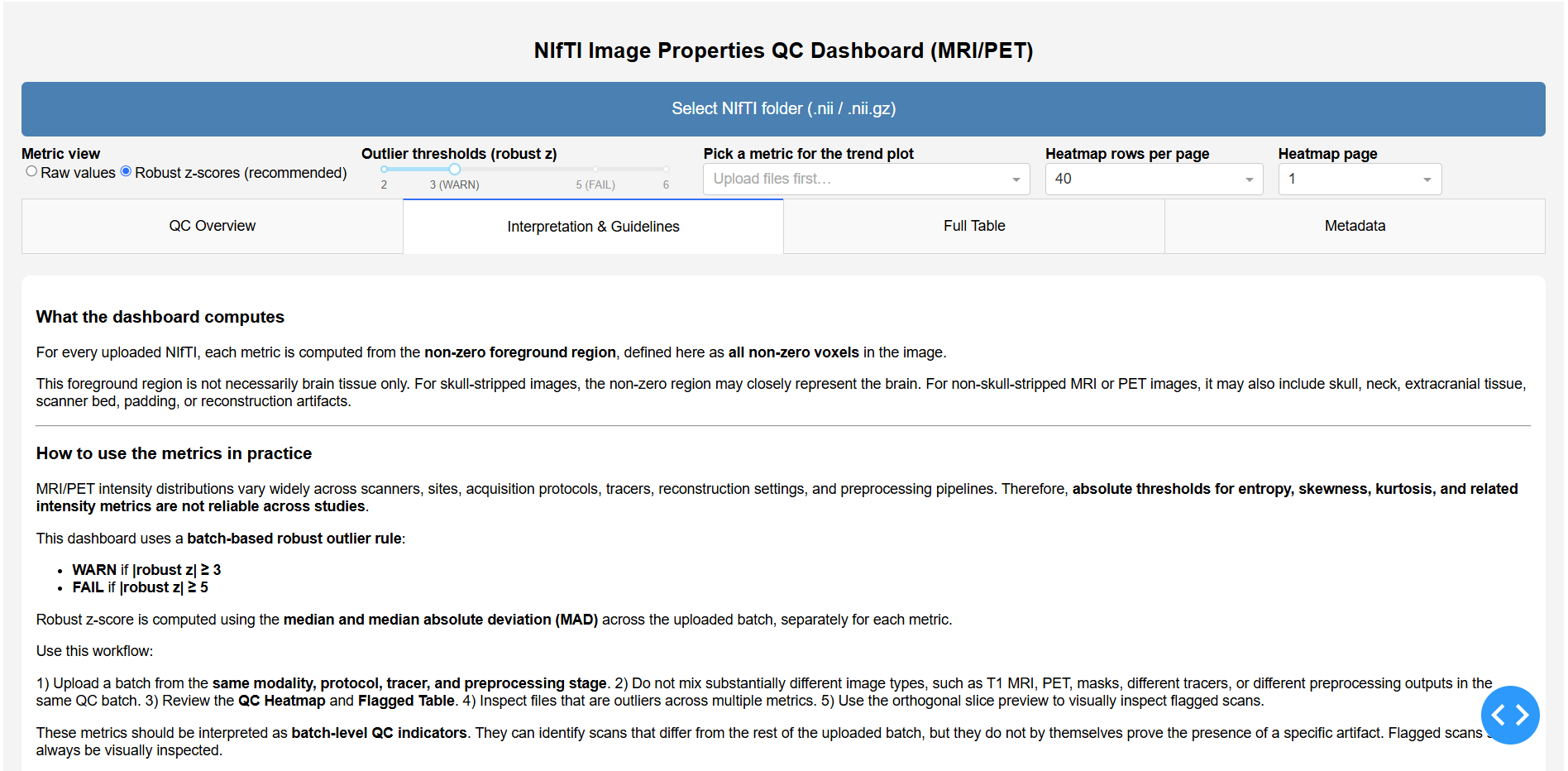}
    \caption{%
    \textbf{Interpretation \& Guidelines tab of the NIfTI Image Properties QC Dashboard.}
    This page provides built-in guidance for interpreting the reported image-property metrics and for using the dashboard in practice. It also summarizes the \textbf{WARN} and \textbf{FAIL} rules based on robust z-scores computed from the batch median and median absolute deviation, outlines a recommended review workflow, and gives practical interpretations of histogram summaries such as entropy, skewness, and kurtosis to support manual follow-up of flagged scans.}
    \label{fig:artifact-dashboard-4}
\end{figure}

Overall, the dashboard is intended to flag scans for review. Scans flagged as WARN or FAIL should be visually inspected before making a final QC decision. The dashboard summarizes these interpretation rules, robust-z thresholds, and recommended review steps in the \textit{Interpretation \& Guidelines} tab (Figure~\ref{fig:artifact-dashboard-4}).

\FloatBarrier

\FloatBarrier

\FloatBarrier

\section{The Dallas Lifespan Brain Study: Quality Control Review of Wave 3 MRI and PET Images}
\label{sec:dlbs-wave3-image-qc-review}

\begin{figure}[htbp]
    \centering
    \includegraphics[width=0.75\linewidth]{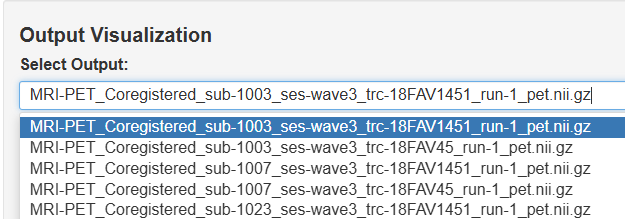}
    \caption{\textbf{Output visualization after batch preprocessing} The dropdown menu lists multiple processed cases, allowing users to select and review each case's outputs.}
    \label{fig:batch_outputs_dropdown}
\end{figure}

Image-level quality control was performed for Wave 3 MRI and PET images from The Dallas Lifespan Brain Study that were included in the external validation dataset. The QC dashboard was applied to the available Wave 3 structural MRI, amyloid PET, and tau PET scans. For each modality, image-property metrics were computed across the full batch, and robust z-scores were calculated using the complete set of scans within the reviewed batch. The heatmap visualization was then used to identify scans with unusual metric values, and selected scans were visually inspected using the axial, sagittal, and coronal preview panels. After batch preprocessing, the output visualization interface allowed users to select individual processed cases from a dropdown menu and review the corresponding outputs (Figure~\ref{fig:batch_outputs_dropdown}).

\begin{figure}[htbp]
\centering
\includegraphics[width=\textwidth]{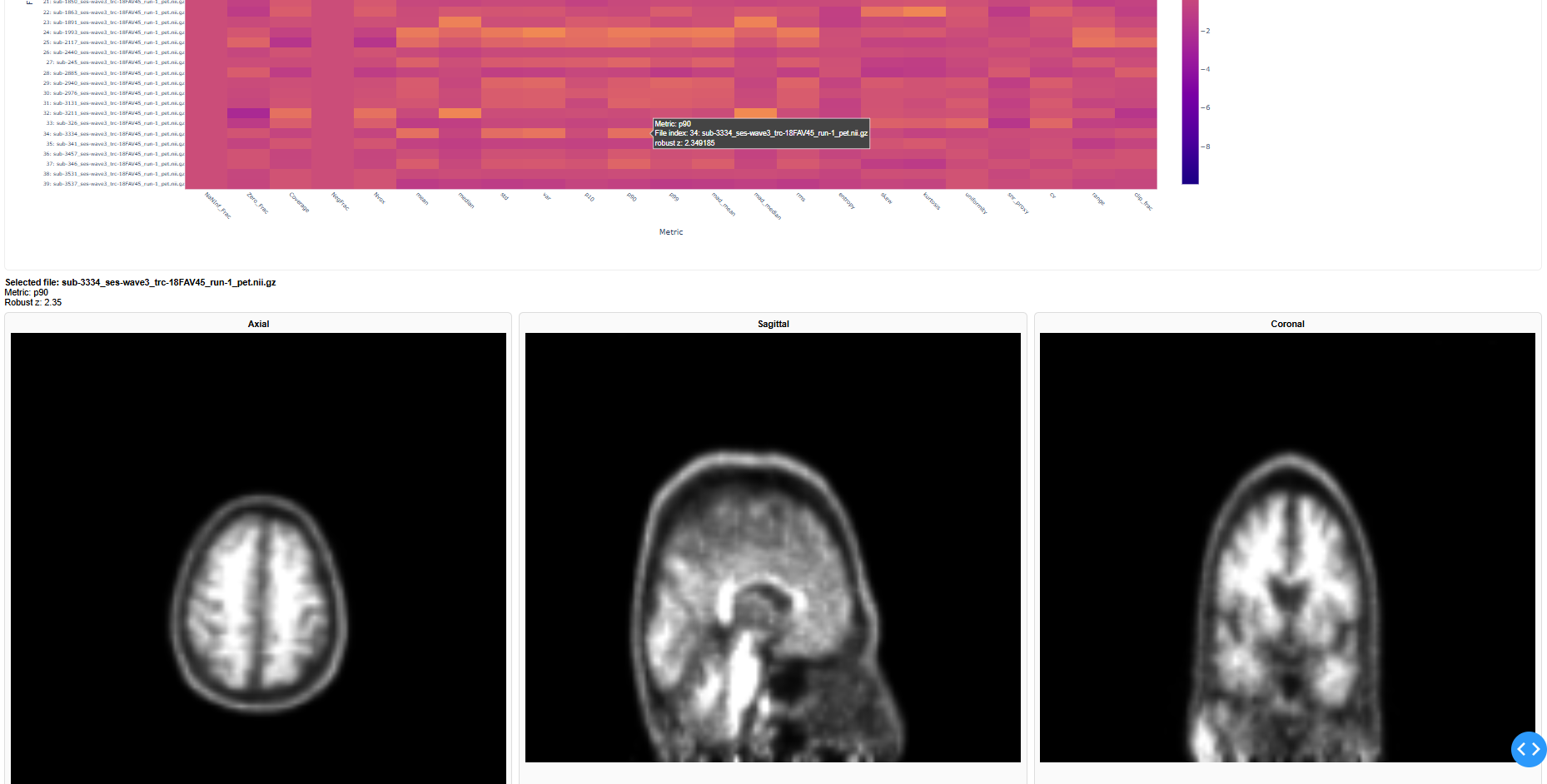}
\caption{\textbf{Representative quality control review of a Wave 3 amyloid PET scan.} The QC heatmap shows robust z-scores for image property metrics, and the orthogonal slice preview shows axial, sagittal, and coronal views of the selected scan. The selected amyloid PET image did not show visible brain image quality concerns.}
\label{fig:amyloid-qc-normal-1}
\end{figure}

\begin{figure}[htbp]
\centering
\includegraphics[width=\textwidth]{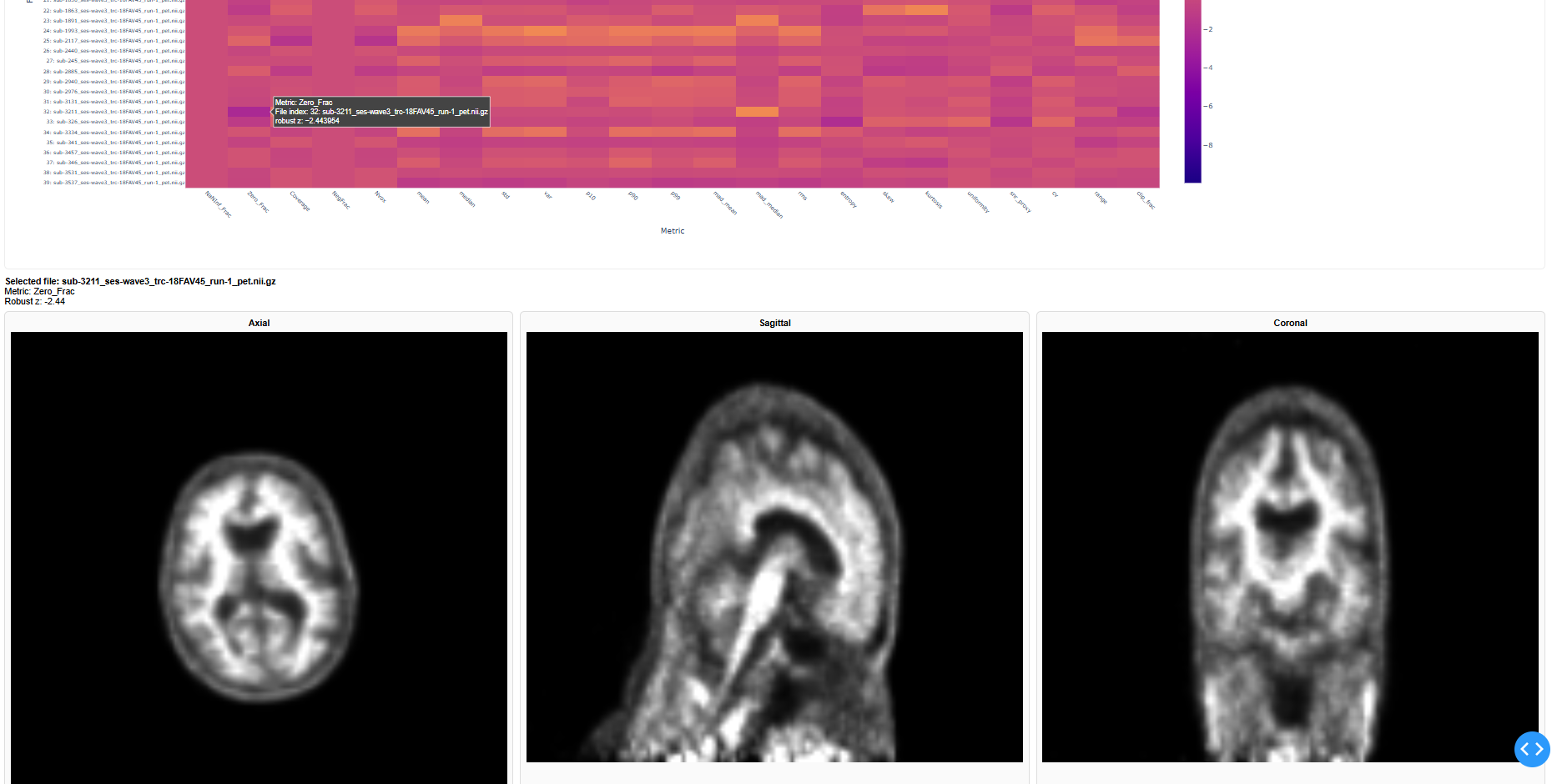}
\caption{\textbf{Additional representative quality control example for a Wave 3 amyloid PET scan.} The selected scan showed expected visual appearance in the orthogonal slice preview, with no obvious artifact affecting the brain region.}
\label{fig:amyloid-qc-normal-2}
\end{figure}

\begin{figure}[htbp]
\centering
\includegraphics[width=\textwidth]{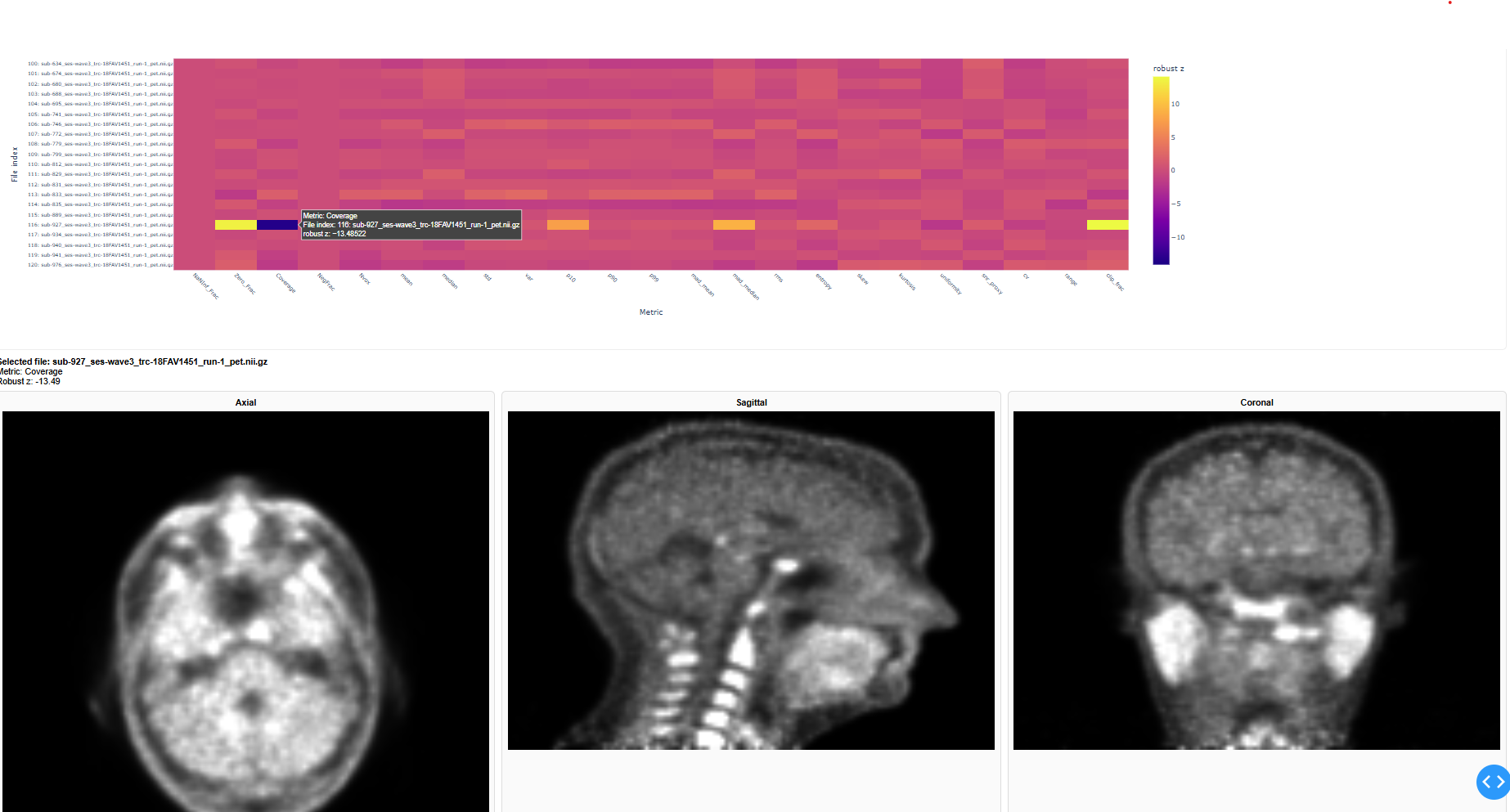}
\caption{\textbf{Quality control review of a flagged Wave 3 tau PET scan.} The scan showed an extreme robust z-score for a coverage related image property metric. Visual inspection indicated that the brain region itself was acceptable and that the flagged metric was primarily driven by greater inclusion of inferior neck/extracranial anatomy rather than a brain image artifact. Based on this review, the scan was retained in the external validation dataset.}
\label{fig:tau-qc-outlier}
\end{figure}

\begin{figure}[htbp]
\centering
\includegraphics[width=\textwidth]{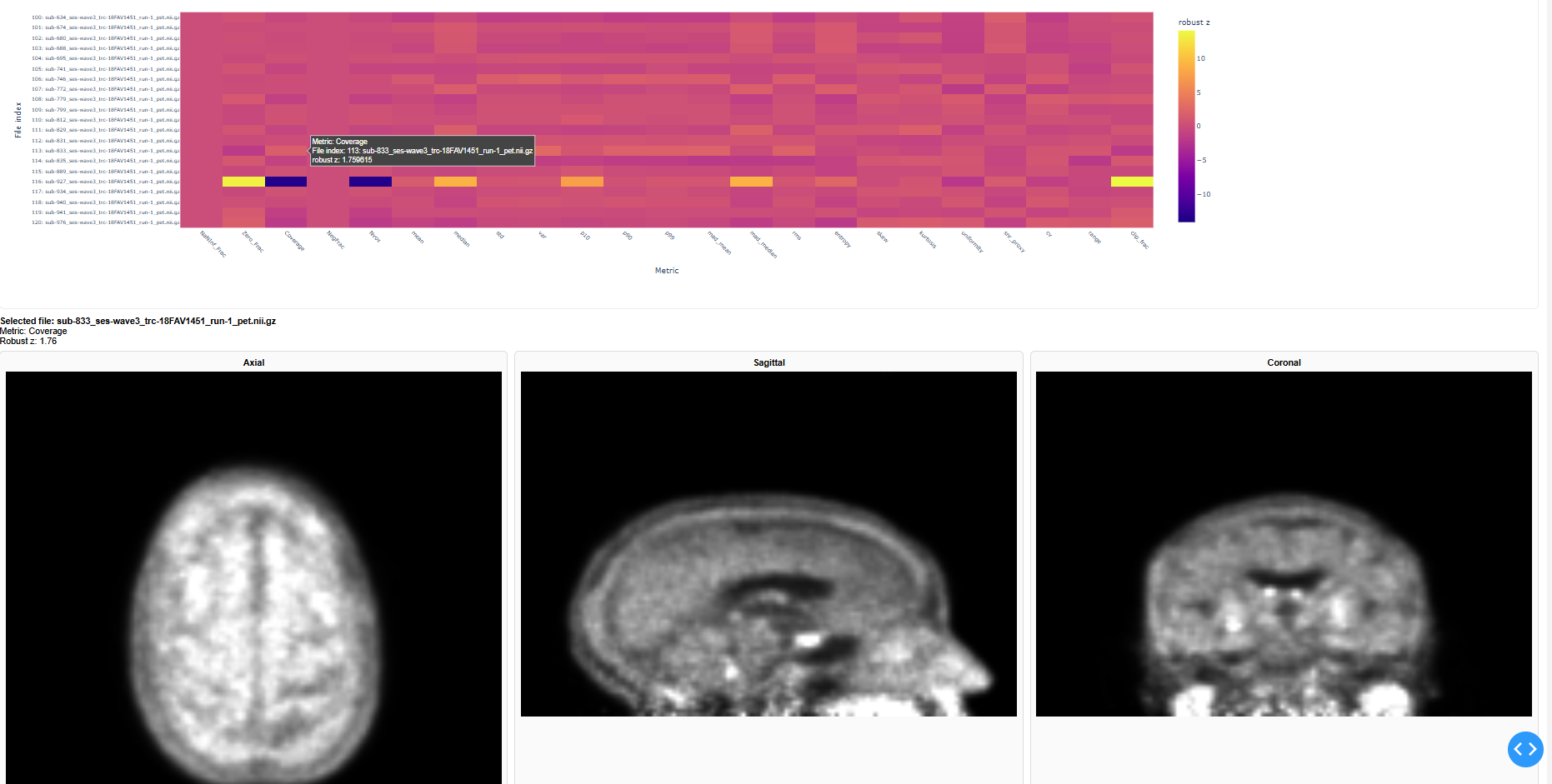}
\caption{\textbf{Additional representative quality control review of a Wave 3 tau PET scan.} The selected scan showed acceptable visual appearance in the axial, sagittal, and coronal views, with no obvious artifact affecting the brain region.}
\label{fig:tau-qc-normal}
\end{figure}

\begin{figure}[htbp]
\centering
\includegraphics[width=\textwidth]{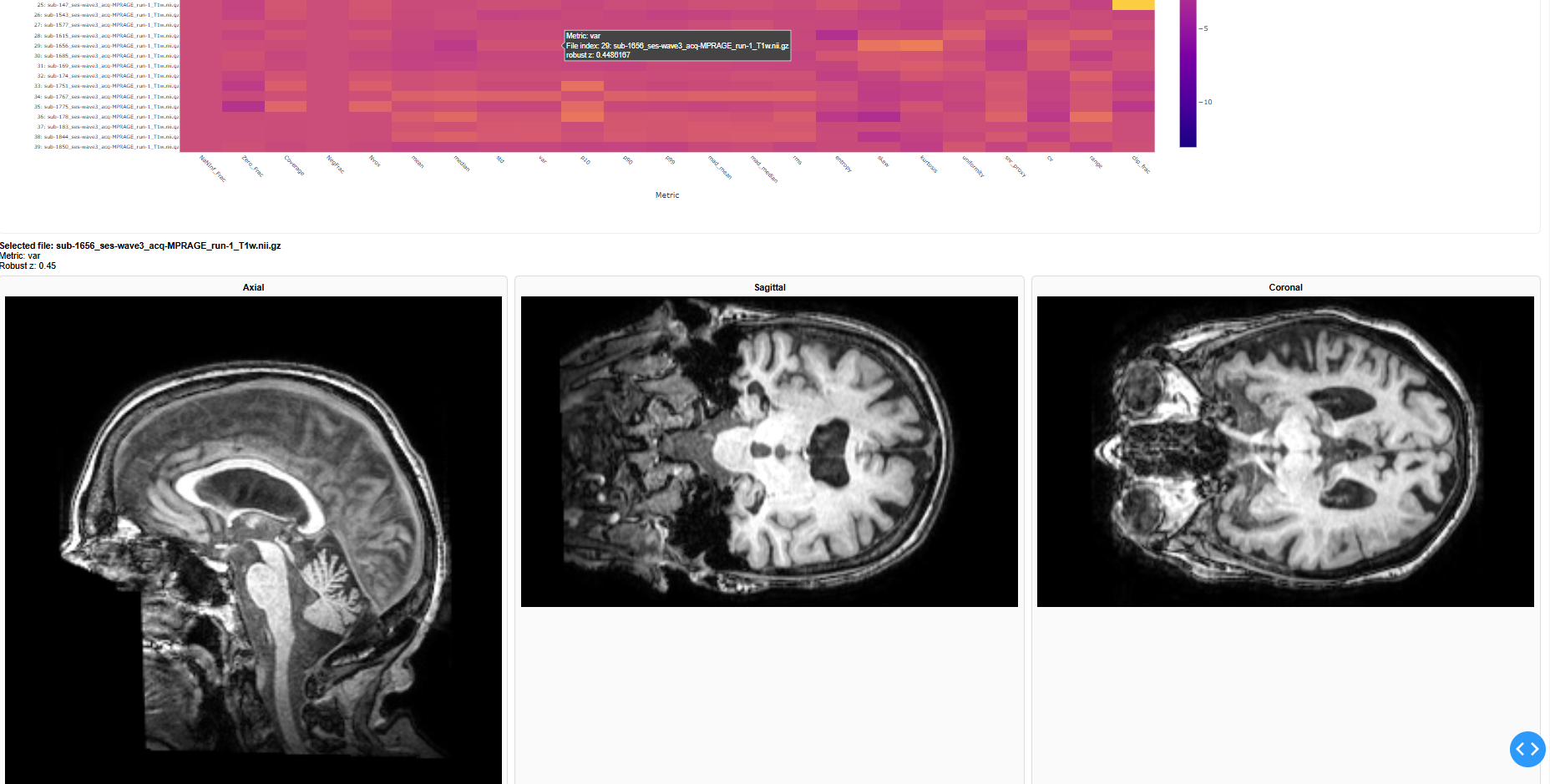}
\caption{\textbf{Representative quality control review of a Wave 3 structural MRI scan.} The dashboard heatmap was used to identify metric level outliers, and the selected MRI was visually reviewed in axial, sagittal, and coronal views. The brain image appeared visually acceptable for downstream processing.}
\label{fig:mri-qc-normal-1}
\end{figure}

\begin{figure}[htbp]
\centering
\includegraphics[width=\textwidth]{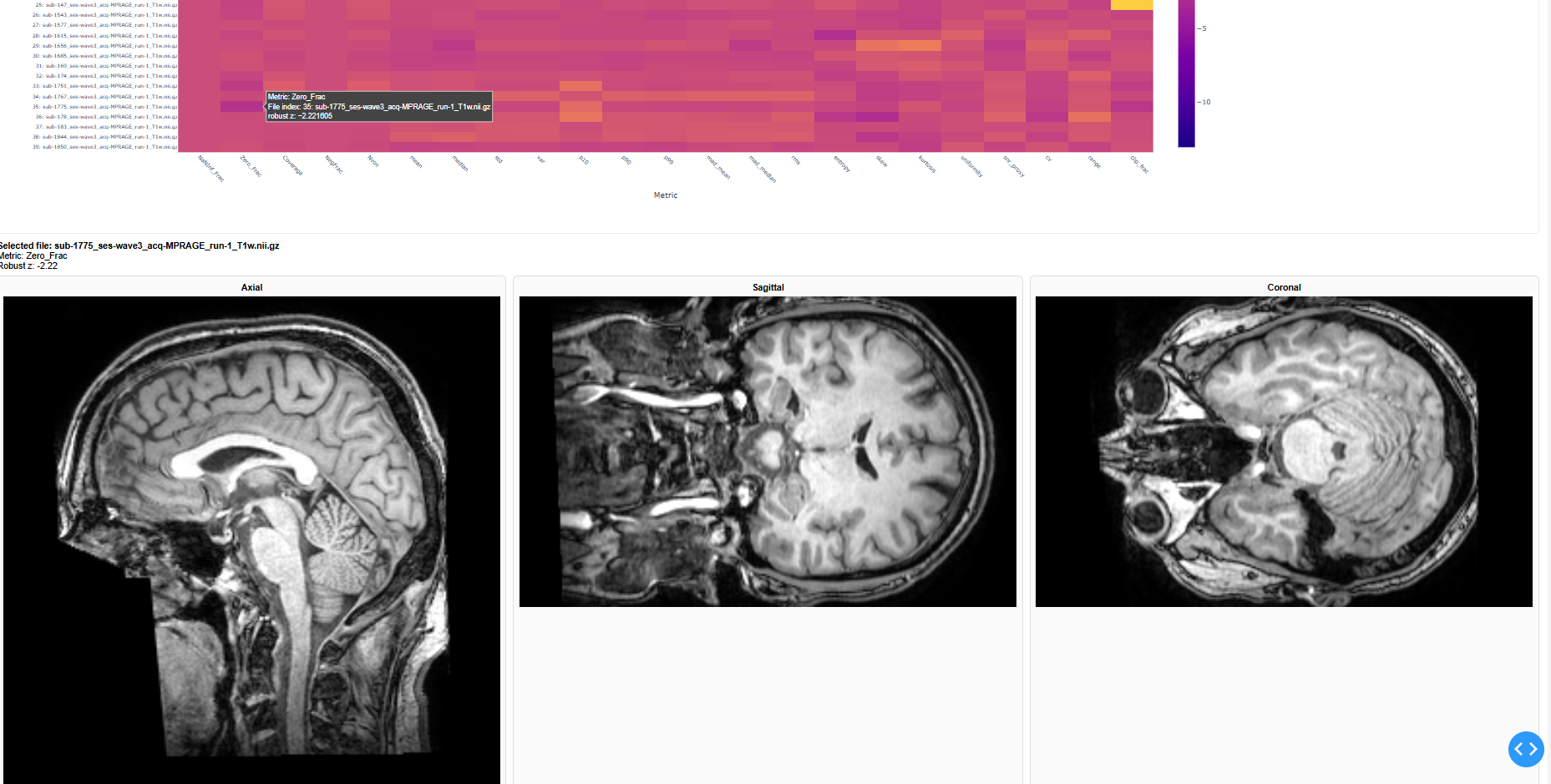}
\caption{\textbf{Additional representative quality control example for a Wave 3 structural MRI scan.} Although image property metrics varied across the batch, the visually inspected MRI showed no visible brain image quality issue in the orthogonal preview and was retained for downstream analysis.}
\label{fig:mri-qc-normal-2}
\end{figure}

Overall, the reviewed Wave 3 MRI and PET images showed acceptable visual quality for downstream external validation. Most scans did not show obvious artifacts affecting the brain region. Representative examples of normal amyloid PET, structural MRI, and tau PET scans are shown in Figures~\ref{fig:amyloid-qc-normal-1}--\ref{fig:mri-qc-normal-2}. One tau PET scan was flagged by the dashboard because of an extreme image property metric related to the field of view/foreground coverage. Visual inspection showed that the brain region itself was acceptable, and the difference appeared to be driven primarily by greater inclusion of inferior neck/extracranial anatomy rather than a brain image artifact. Therefore, this scan was retained in the validation dataset.

\FloatBarrier

\FloatBarrier

\FloatBarrier

\FloatBarrier

\FloatBarrier

\FloatBarrier

\FloatBarrier

\subsection{Post-registration QC of TRAECR outputs}

Post-registration QC was also performed on the registered PET outputs generated by the TRAECR workflow. The registered amyloid and tau PET images were evaluated using the QC dashboard to determine whether image property metrics could help identify problematic registrations after pre-processing. The coverage metric was particularly informative: visibly failed registrations often appeared as coverage outliers, whereas visually successful registrations generally remained near the expected batch distribution. For example, an amyloid PET output with visible registration failure showed a coverage robust z-score of $-3.12$, and a tau PET output with visible registration failure showed a coverage robust z-score of $-7.16$. In contrast, visually successful amyloid and tau PET registrations showed coverage robust z-scores of $0.21$ and $0.48$, respectively. These examples indicate that post-registration QC using the dashboard can help prioritize outputs requiring closer review, with visual inspection used as the final confirmation step.

\begin{figure}[htbp]
\centering
\includegraphics[width=\textwidth]{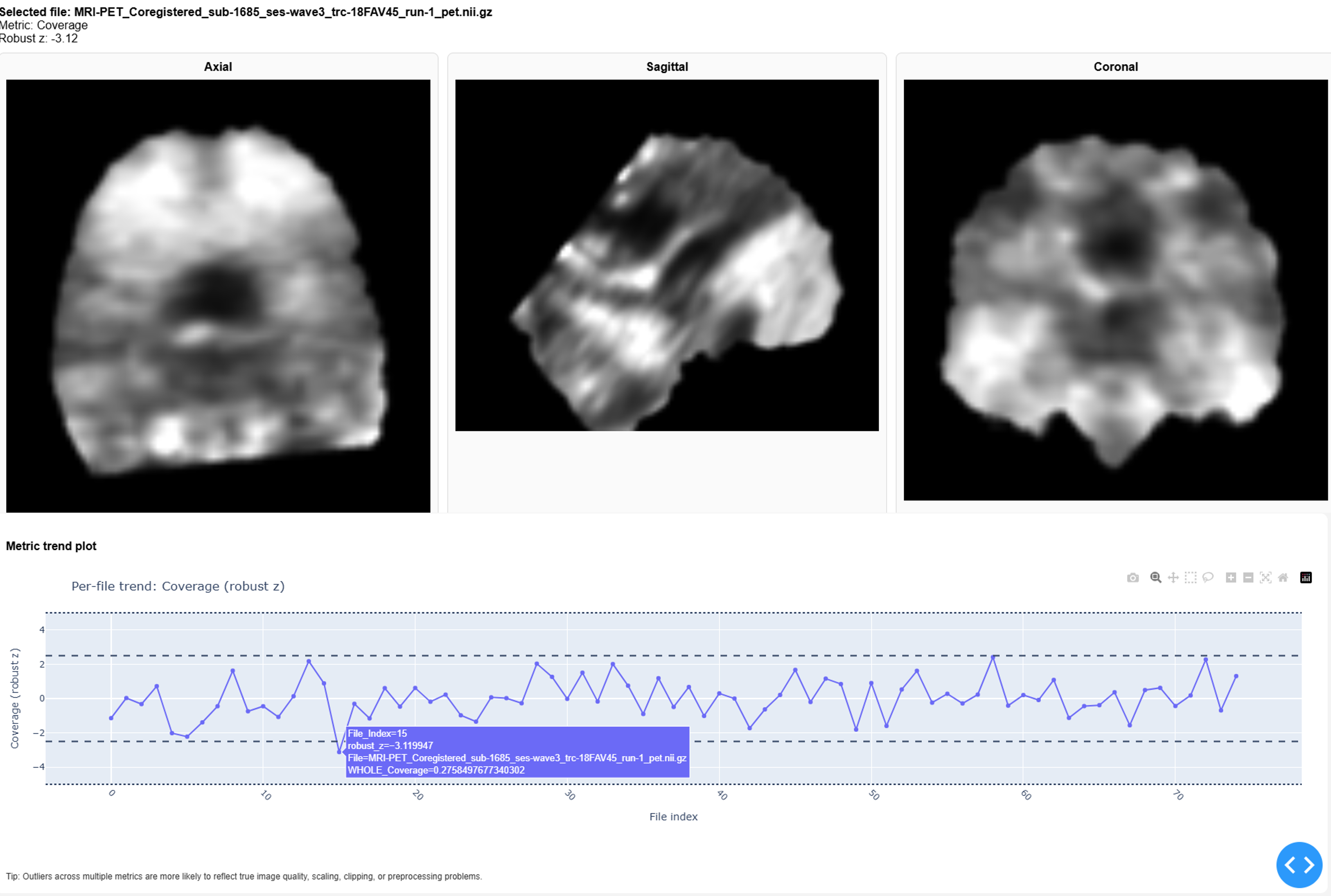}
\caption{\textbf{Post-registration QC example for a failed amyloid PET registration.} The registered amyloid PET output showed visible anatomical misalignment in the orthogonal slice views. The coverage metric was flagged as a WARN-level outlier with a robust z-score of $-3.12$, indicating abnormal foreground coverage relative to the registered amyloid PET batch.}
\label{fig:postreg-qc-amyloid-failed}
\end{figure}
\FloatBarrier

\begin{figure}[htbp]
\centering
\includegraphics[width=\textwidth]{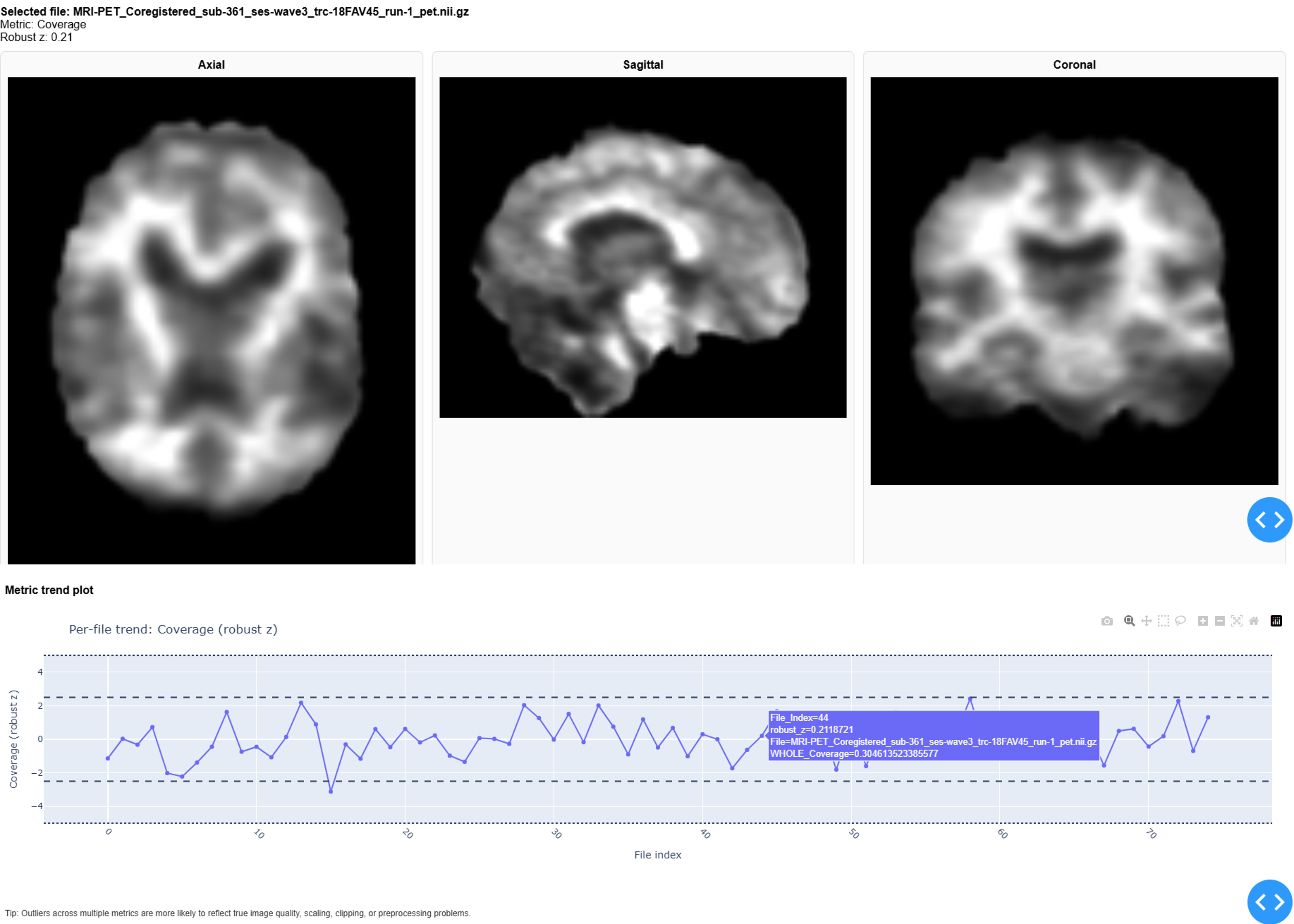}
\caption{\textbf{Post-registration QC example for a successful amyloid PET registration.} The registered amyloid PET output showed acceptable anatomical alignment in the axial, sagittal, and coronal views. The coverage robust z-score was $0.21$, remaining within the expected batch range.}
\label{fig:postreg-qc-amyloid-successful}
\end{figure}
\FloatBarrier

\begin{figure}[htbp]
\centering
\includegraphics[width=\textwidth]{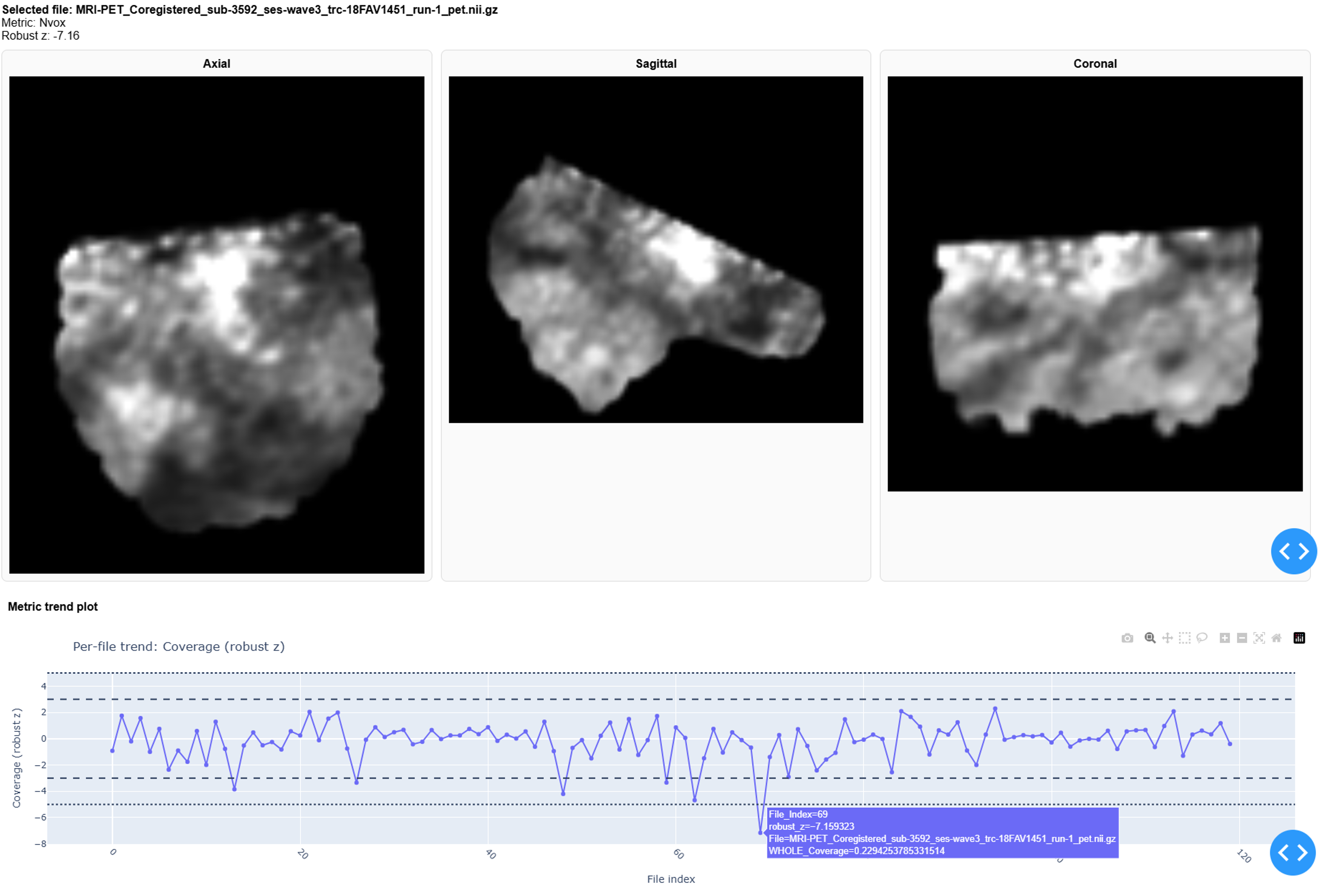}
\caption{\textbf{Post-registration QC example for a failed tau PET registration.} The registered tau PET output showed visible anatomical misalignment in the orthogonal slice views. The coverage metric was flagged as a FAIL-level outlier with a robust z-score of $-7.16$, indicating markedly abnormal foreground coverage relative to the registered tau PET batch.}
\label{fig:postreg-qc-tau-failed}
\end{figure}
\FloatBarrier

\begin{figure}[htbp]
\centering
\includegraphics[width=\textwidth]{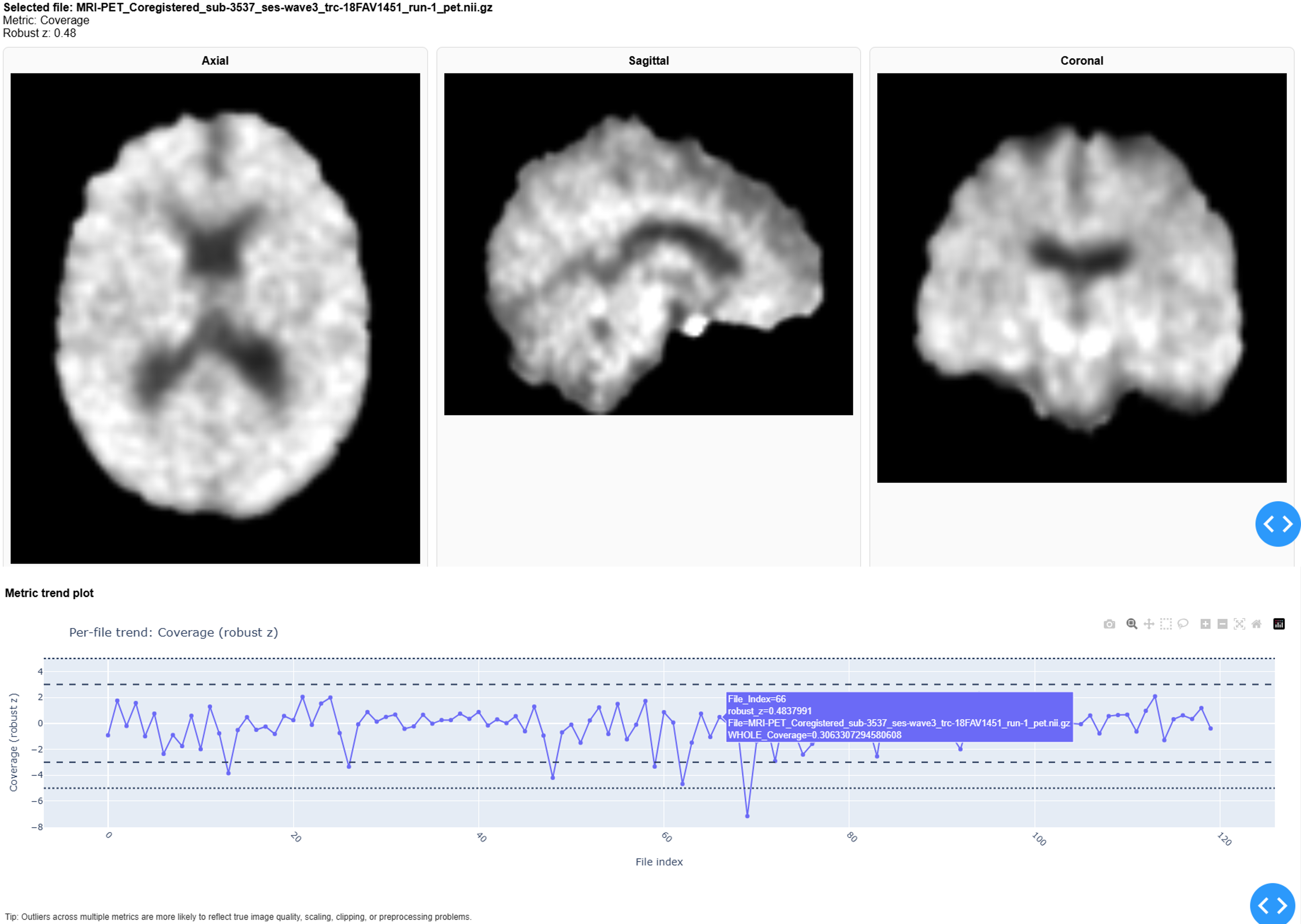}
\caption{\textbf{Post-registration QC example for a successful tau PET registration.} The registered tau PET output showed acceptable anatomical alignment in the axial, sagittal, and coronal views. The coverage robust z-score was $0.48$, remaining within the expected batch range.}
\label{fig:postreg-qc-tau-successful}
\end{figure}
\FloatBarrier